\documentclass[preprint2]{aastex}

\shorttitle{COMPARATIVE STUDY OF ASYMMETRY ORIGIN OF GALAXIES IN DIFFERENT ENVIRONMENTS}
\shortauthors{Plauchu-Frayn \& Coziol}

\begin{document}

\title{Comparative Study of Asymmetry Origin of Galaxies in \\ Different Environments. 
II. Near-Infrared observations.}

\author{I. Plauchu-Frayn \altaffilmark{1} \& R. Coziol \altaffilmark{1}}
\email{plauchuf@astro.ugto.mx, rcoziol@astro.ugto.mx}
\altaffiltext{1}{Departamento de Astronom\'{\i}a, Universidad de Guanajuato Apartado Postal 144, 36000 \\Guanajuato, Gto, M\'exico}

\begin{abstract}

In this second paper of two analysis, we present near-infrared 
morphological and asymmetry studies perfomed in sample of 92 
galaxies found in different density environments: galaxies in 
Compact Groups (HCGs), Isolated Pairs of Galaxies (KPGs), and 
Isolated Galaxies (KIGs). Both studies have proved useful 
to identify the effect of interactions on galaxies.

In the NIR, the properties of the galaxies in HCGs, KPGs and KIGs 
are more similar than in the optical. This is because the NIR band 
traces the older stellar populations, which formed earlier and are 
more relaxed than the younger populations. However, we found 
asymmetries related to interacions in both, KPG and HCG samples. 
In HCGs, the fraction of asymmetric galaxies is even higher than 
what we found in the optical.

In the KPGs, the interactions look like very recent events, while in
the HCGs, galaxies are more morphologically evolved and show 
properties suggesting they suffered more frequent interactions. 
The key difference seems to be the absence of star formation 
in the HCGs: while interactions produce intense star formation 
in the KPGs we do not see this effect in the HCGs. This is consistent 
with the dry merger hypothesis \citep{cp07}: 
the interaction between galaxies in compact groups, (CG), are happening 
without the presence of gas. If the gas was spent in stellar 
formation (to build the bulge of the numerous early-type galaxies), 
then the HCGs possibly started interacting sometime before the KPGs. 
On the other hand, the dry interaction condition in CGs suggests 
the galaxies are on merging orbits, and consequently such system 
cannot be that much older either. Cosmologically speaking, the 
difference in formation time between pairs of galaxies and CGs may 
be relatively small. The two phenomena are typical of the formation 
of structures in low density environments. Their formation represent 
relatively recent events.

\end{abstract}

\keywords{galaxies: interactions -- galaxies: photometry -- galaxies: structure}

\section{Introduction}

This paper presents the second of two analyses about the importance 
of environment on the formation and evolution of galaxies observed in 
the nearby universe. In the first paper (Plauchu-Frayn
\& Coziol 2010; hereafter Paper~I), we have presented the results of
our morphological and asymmetry study in the optical ($V$ and $I$
filters) for a sample of 214 galaxies found in three different density 
environments:  galaxies in Compact Groups (HCGs), Isolated Pairs of
Galaxies (KPGs),  and Isolated Galaxies (KIGs) \citep{kara72, kara73, hickson82}. 
We have also performed a comparative statistical analysis of the isophotal 
and asymmetry properties of these galaxies.

In the optical, we observed some clear differences in the properties of
galaxies in the different environments. First, isolated galaxies
tend to be more compact and symmetrical than galaxies in pairs or in
compact groups (CGs). It suggests interactions produce stellar
orbits with higher energies. Second, evidence for interactions seems
more obvious for galaxies in pairs than for galaxies in CGs. Because
the HCGs are richer in early-type galaxies than the KPGs, these
differences suggest interactions played a more important role in the
CGs in the past. Either because the high density environment of CGs 
favors more interactions and consequently the galaxies evolve more rapidly 
in such systems, or because galaxies in CGs started interacting earlier 
in the past than those in pairs.

The present paper extends our morphological and asymmetry study to
the near-infrared, (NIR), using deep $J$ and $K'$ images. Contrary to the
optical, the NIR bands are sensible to the older, less massive but
dominant stellar populations and trace better the distribution of
the mass of galaxies \citep{frogel78}. Consequently, using NIR images 
allow us to compare interaction effects over different time scales than 
in the optical. This is important in CGs, where it is suspected that 
galaxies  interact and merge under dry conditions \citep{cp07}.

Our analysis is similar to the one used in the optical (Paper~I). We
have applied independently two different methods: fitting of
elliptical ellipses on the isophotal levels of the galaxies and 
determination of their asymmetry level. In Section~2, we present the
properties and selection of the sub-samples observed in the
NIR; in Section~3, we describe the conditions of observations and
explain the reduction process; in Section~4, we present the results
of our photometry and asymmetry analyses, and the comparative
statistical studies in different environments; in Section~5, we
compare the level of nuclear activity star formation or active 
galactic nucleus, (AGN) in the different environments using spectra 
extracted from SDSS\footnote{{\it Sloan Digital Sky  Survey} http://www.sdss.org}; 
in Section~6, we discuss our results, comparing them with those 
obtained in the optical; our conclusions are stated in Section~7.

\section{Selection and properties of the observed galaxies}

The sub-sample for our NIR analysis consists of 92 galaxies taken
from our list of galaxies previously observed in the optical. A
detailed description of the properties of these galaxies can be
found in Paper~I. Only 21 galaxies were part of our previous NIR
analysis \citep{cp07}. The remaining 71 galaxies are new observations.

The galaxies observed in NIR form only 50\% of those observed in the
optical. This is because observations in NIR are more complicated
and time expensive than in the optical. In particular, in the NIR,
one has to move the telescope constantly forming a mosaic sequence.
This technique is used to avoid ghost images (remnants of galaxy
image on the pixels if the target is kept at the same position) and
to allow proper sky subtraction. In setting up our sub-sample,
therefore, a supplementary step consisted in selecting galaxies with
a semi-major axis in the range 15''$\le a \le$ 90'', to allow the
telescope to move in a cross or a square pattern.

The properties of the galaxies observed are reported in
Table~\ref{tabl1} for the KIGs, Table~\ref{tabl2} for the KPGs, and
Table~\ref{tabl3} for the HCGs. As explained in Paper~I, morphological 
types for the KIG sample were taken from Sulentic et al. (2006). For
the KPG and HCG samples these were determined based on our
own CCD images. Evidence for bars was added based on the optical
images and/or our isophotal study. No new bars were added based on the
NIR images.

\begin{figure*} 
\epsscale{1.0} \plotone{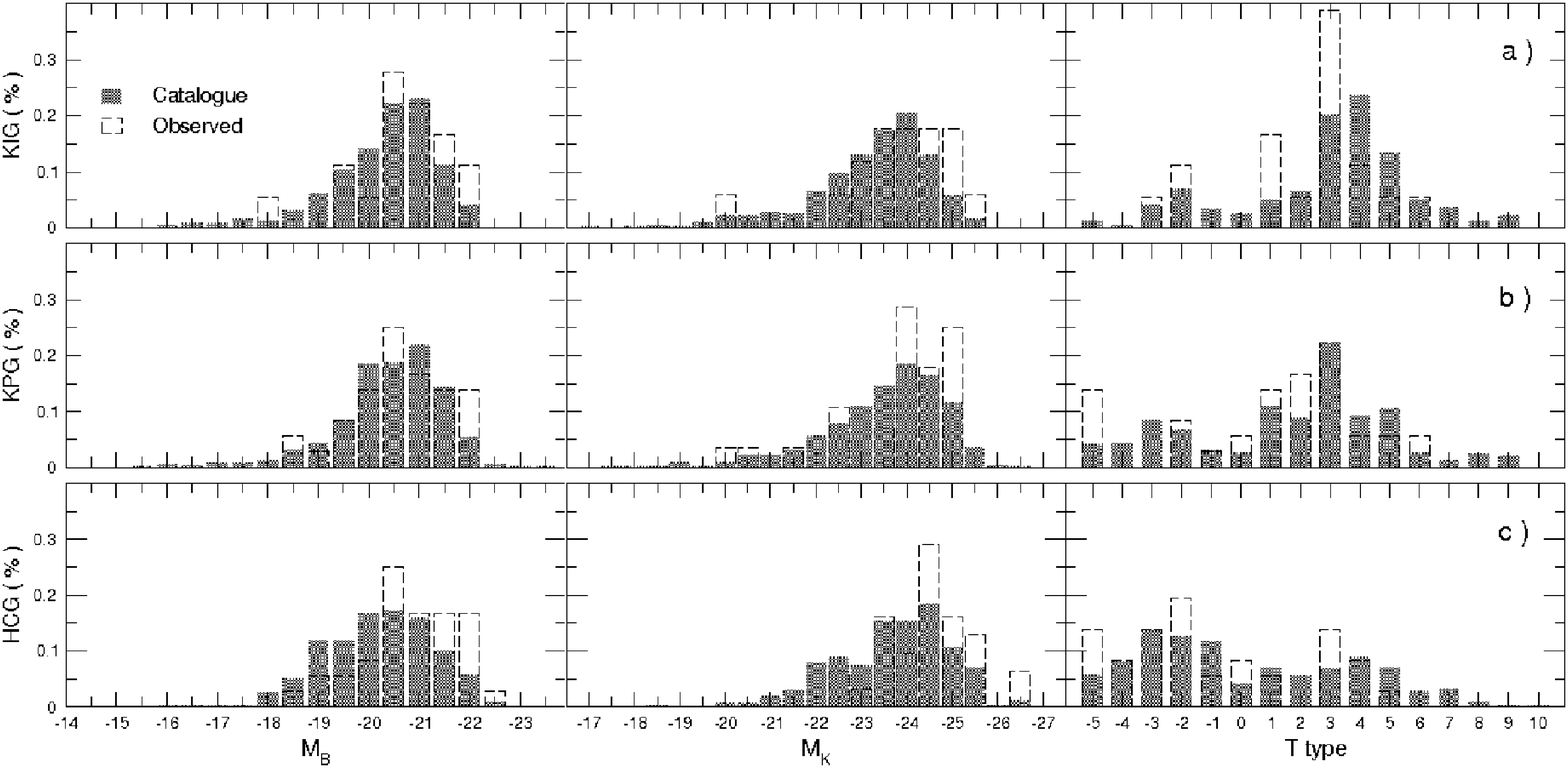} 
\caption{ Distribution of catalog
versus observed galaxies: (a) KIG, (b) KPG, and (c) HCG galaxies.
\label{fig1}}
\end{figure*}

In Figure~\ref{fig1}, we compare the properties of the observed
samples in the NIR with the properties of the galaxies in their
respective catalogs. The $B$ and $K_S$ magnitudes \citep{paturel94, 
paturel97, jarret00} and revised numerical code of the
morphological type \citep{vauc91} were taken from the
HyperLeda\footnote{{\it HyperLeda database} http://leda.univ-lyon1.fr} and 
2MASS\footnote{{\it 2 Micron All Sky Survey} http://irsa.ipac.caltech.edu} 
databases.  Also, in Table~\ref{tabl4} are compared the diameters in 
both magnitudes  and the redshifts, as found in the HyperLeda and 2MASS 
databases. One can see that the observed samples are fairly representative 
of their parent samples. In Table~\ref{tabl4}, the results of the Mann-Whitney 
statistical tests are consistent with no difference in absolute magnitude for
the observed KPGs. In the case of the KIGs, the observed sample is
slightly more luminous than in the whole catalog. This is because
they are located slightly nearer than the galaxies in the whole
catalog. The HCGs are also brighter and bigger in the observed 
sample than in the whole catalog. These differences are
consequence of the criteria used in preparing the NIR observations
and will be taken into account during our analysis.

\section{Observation and reduction}

We have obtained new NIR images in $J$ (1.28 $\mu$m) and $K'$ (2.12
$\mu$m) bands for 71 galaxies with the 2.1m telescope of the Observatorio
Astron\'omico Nacional, located in the Sierra San Pedro M\'artir, in
Baja California, M\'exico. Five different runs were necessary
(Table~\ref{tabl5}). The instrument used was CAMILA \citep{cruz94}. 
This camera is formed of four NICMOS3 detectors with
256$\times$256 pixels, which are sensitive over the range 1 to 2.5
$\mu m$. This instrument includes a diaphragm with a wheel of
filters, which are cooled to reduce the background radiation level.
The optical system, which consists of a mirror and a focal
reducer designed for the f/4.5 secondary, gives a 3.6$'\times$3.6$'$
field of view, corresponding to a plate scale of 0.85$''$pixel$^{-1}$.

Because the detector in the NIR saturates rapidly (due to the
brightness of the sky), several short exposures were obtained and
combined to form one final image. For each filter we have used the
longest integration time possible that keeps the total counts within
the linear range of the detector response. For the $J$ filter, we took
for each galaxy $\sim$40 images of 80 seconds. For the $K'$ filter,
we took $\sim$80 images of $\sim$15 seconds.

As mentioned in the previous section, a mosaic technique was used in
order to map properly sample the quickly varying sky background. The
technique consists in moving the telescope between each short
exposure. The way the telescope moves depends on the angular size of
the target galaxy. Usually we form a cross or a square, but sometimes
we must alternate between the target and the sky when the object
covers the whole detector (for example, the whole CG). In general,
the combination of a sequence of mosaic images yields a total
integration time which is equal or below the observation time. This is
due to the rejection of some bad images in a sequence. In
Tables~\ref{tabl1}$-$\ref{tabl3}, the columns~6 and
7 list the final total exposure time in $J$ and K$'$, respectively,
for each galaxy. Note that to increase the S/N and/or to replace bad
images, many galaxies were observed more than one night, but only
during the same observing run.

The nights were clear and photometric. Usually, and as expected for
NIR, the observations are not affected on full moon nights. However, in a 
few cases, where the moonlight was reflected by the dome  we have
discarded the images.

For photometry calibration several standard stars were observed
during each observing run. These stars were taken from the UKIRT\footnote{
The United Kingdom Infrared Telescope is operated by the joint Astronomy
Centre on behalf of the Science a Technology Facilities Council of U.K.} 
\citep{hawarden01} extended standards list, available at the 
WEB page of the observatory\footnote{
www.astrossp.unam.mx/estandar/standards/fs$\_$extended.html}. The
instrumental magnitudes for the standard stars were estimated by
measuring the flux of each observed star after correcting for the
atmospheric extinction. The calibration equations were calculated by
fitting linear regressions on the observed values. For the
photometric error we adopted the standard deviation between our
estimated magnitude and the magnitude determined in the UKIRT 
standards list. Magnitudes for the observed galaxies have also been
corrected for galactic extinction \citep{schlegel98}. Because
the galaxies observed are at $z \le0.04$, no $K$-correction has been
applied. Details for the observations are given in Table~\ref{tabl5}. The
calibration in flux was applied only after the different analyses
(isophotal and asymmetry) were performed. This way of doing keeps the
S/N in the different images as high as possible.

The reduction and calibration processes were standard and within
IRAF\footnote{ IRAF is the Image  Analysis and Reduction Facility
made available to the astronomical community by the National Optical
Astronomy Observatory, which is operated AURA, Inc., under
contract with U.S. National Science Foundation.}. One useful
characteristic of CAMILA is that it subtracts a bias image from the
target image after each exposure. This reduction step was
consequently not needed. The next step consisted in trimming the
individual images. This is done to remove the various bad lines and
columns at the edges of the detector and to eliminate the effect due
to vignetting. We subsequently applied a mask on all the images to
remove the bad pixels. Normalized sky-flats in each filter were used to
correct for the differences in quantum response. These normalized
flats were obtained by fitting a 2D surface (with IMSURFIT) on
combined sky-flats.

In the NIR, one important step of the reduction process is eliminate
the sky contribution. To do so, we formed a median sky image by
combining four  adjacent exposures in the mosaic sequence and
subtracted it from the corresponding exposure in the sequence. After
aligning and trimming the images, the sequence of exposures was
combined. First, the $(x,y)$ coordinates of the target in each image
were measured using the circular aperture photometry option. These
positions were used to determine the optimal shift needed to put the
target in the different frames at exactly the same position, keeping
the field of view as large as possible. These shifts were of the
order of a tenth of a pixel. After shifting, we trimmed the individual
images to the most optimum size. Before combining the
exposures, most of the cosmic rays were removed using $COSMICRAYS$
task. In some cases it was necessary remove by hand the remaining cosmic 
rays  by using the $IMEDIT$ task. An average image was then formed by
combining the whole sequence of exposures.

\section{Results of analysis}

It is known, that during interactions some of the properties of galaxies
are affected. Properties such as morphology, brightness, color and nuclear
activity are changed as a result of the disruption of gas and stellar component 
during the encounter \citep{larson78, kennicutt84, lui95, laurikainen00, patton05, 
woods07, mcintosh08}.

To study further this, for each galaxy two different analysis were performed: 
fitting of ellipses on the NIR isophotes and estimation of the level of
asymmetry. The application of these methods is similar to that used in
the optical. A detailed explanation of the methods can be found in 
Paper~I. For homogeneity sake, the analysis for the 21 HCG galaxies 
previously observed by Coziol \& Plauchu-Frayn (2007) was completely 
redone. The results of our analysis are reported for individual galaxies 
in Tables~\ref{tabl6}$-$\ref{tabl8} for galaxies in KIG, KPG and HCG samples,
respectively.

By measuring these parameters at the inner and  outer parts in the observed 
galaxies, we plan to study the effect interactions on galaxies in the three 
different environments. To do this, we have chosen $ r_{0}$ to be a radius
independent on the distribution of light inside and outside of which we will 
measure isophotal and asymmetry parameters, following the methods described 
in Paper~I. Using the major axis at 20 mag arcsec$^{-2}$ in $K_s$ band \citep{jarret00} 
as given in 2MASS, we have determined the linear diameters in 
kiloparsec, $D_{K_s}$, for galaxies in the HCG, KPG and KIG catalogs and 
determined the median of the diameters distribution. The median value obtained 
is 14 kpc. In our sample, a few galaxies (22\% of the sample: 8 HCGs, 9 KPGs, 
and 3 KIGs) turned out to have a $D_{K_s}$ which is smaller than this value. 
Consequently, we have used two different $r_{0}$, equal to 3.5 kpc 
(approximately $D_{K_s}/4$) for the normal size galaxies (approximately $M_J <-22$) 
and half this value, 1.8 kpc, for the galaxies with smaller diameters ($M_J > -22$).

Based on the surface brightness profiles, $\mu(r)$ of the galaxies, we have 
estimated the concentration index, $C$. This is defined as (Paper~I): 
$C_{< r_{0}}= \mu_{r_{0}}- \mu_{< r_{0}}$  and $C_{> r_{0}}=\mu_{> r_{0}}-\mu_{r_{0}}$, 
inside and outside $ r_{0}$, respectively. $C$ is a measure of the light concentration 
of a galaxy profile, having high values for centrally concentrated light profiles. 
Early-type galaxies tend to have the most concentrate  light profiles, while 
late-type galaxies have the least concentrated ones \citep{abraham94, shimasaku01}. 
On the other hand,  interactions are expected to perturb the stellar material, 
changing in the process light profiles of galaxies and affecting their 
concentration indices.

Finally, to make our interpretation of the asymmetry study more straightforward, 
we have used a slightly different measure of asymmetry  than that found in 
the literature \citep{schade95, conselice00, hutchings08}. The level of asymmetry 
as a function of the semi-major axis $a$ is estimated by the following formula:

\begin{equation}
A(a)_{180^\circ}\equiv \frac{I_0}{I_{180^\circ}}
\end{equation}

where $I(a)_0$ is the intensity in the original image and
$I(a)_{180}$ is the intensity in the rotated image. This formula yields
values between 1 (completely symmetric) and $>1$\ (completely
asymmetric). We refer the reader to Paper~I for a detailed 
description of this method.

\begin{figure*}
\epsscale{0.8} \plotone{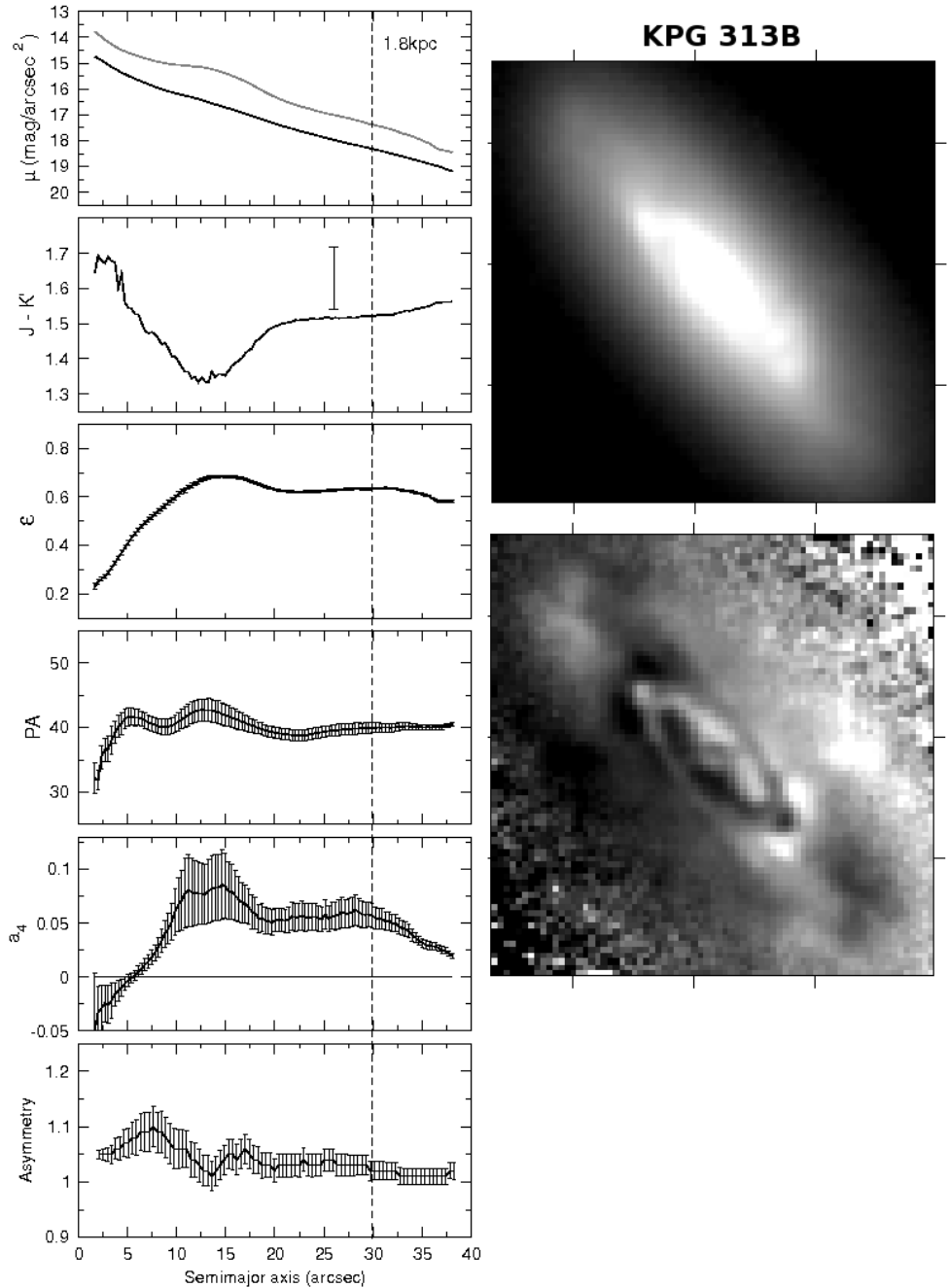} 
\caption{ KPG 313B mosaic. {\it Left
panel}: isophotal parameter profiles as a function of the semi-major
axis $a$ -- from top to bottom: surface brightness in $J$ and $K'$
(mag arcsec$^{-2}$), $J-K'$ color index (magnitude), ellipticity,
position angle (degrees), isophotal deviation from pure ellipse, and
asymmetry level. The dashed vertical line indicates the average half
radius $r_0 = 3.5$ kpc (or $r_0 = 1.8$ kpc when indicated); {\it
Right panel}: J-band image, displayed on a logarithmic scale and the
residual image (bottom image). North is up and east is to the left.
The complete figure set (92 images) is available in the online journal.\label{fig2}}
\end{figure*}

In Figure~\ref{fig2}, we show, as an example, the mosaic for one
very asymmetric galaxy (the full sample of mosaic images is available
in the online version of the journal). On the left of Figure~\ref{fig2}, 
we present the isophotal profiles. The dashed vertical line marks the 
location of the half-radius, $r_{0}$, adopted. On the right of the same 
figure, we present the $J$-band image (or $K'$-band for galaxies that were 
observed only in this filter), displayed on a logarithmic scale. 
We also present the residual image from the asymmetry analysis 
(left bottom image). In all these images, the north is at the top and 
east is to the left.

\subsection{Comparison of galaxies with same morphologies in 
different environments}

We have divided our samples in three morphology groups: early-type
(E--S0), intermediate-type (Sa--Sb), and late-type (Sbc--Im). The median 
values of the properties measured on the observed galaxies in the three 
different groups are reported in Tables~\ref{tabl9}$-$\ref{tabl11} for
early, intermediate and late types, respectively. We now discuss the 
variations on the properties encountered of each group depending on 
their environment. To check for the statistical significance of the 
observed variations, nonparametrical tests (Kruskal-Wallis for three 
samples or Mann-Whitney in case of only two samples) were also performed. 
All the tests were done at a level of significance of 95\%, which is 
the standard for these kind of tests. The results for the statistical 
tests are reported in Tables~\ref{tabl9}$-$\ref{tabl11}.

\subsubsection{Early-type (E--S0) galaxies}

\begin{figure*}
\epsscale{0.7} \plotone{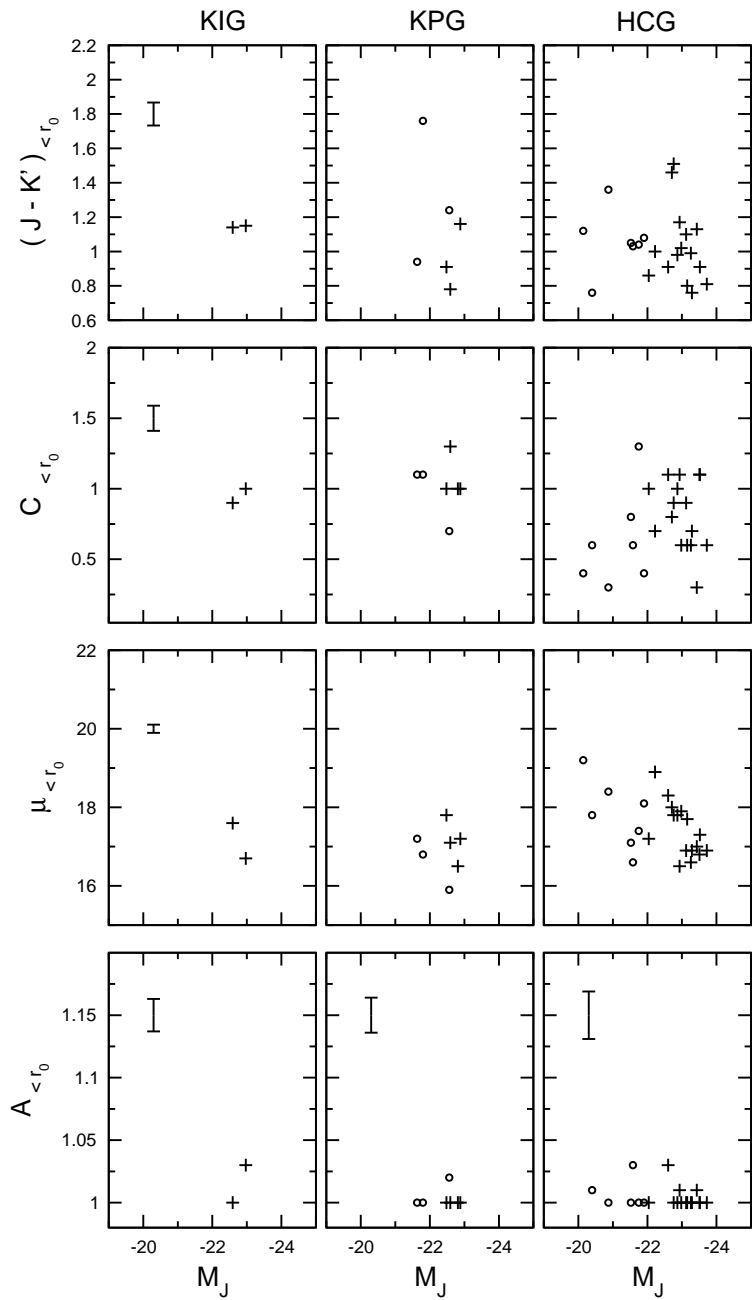}
\caption{Variations in
Early-type galaxies of isophotal parameters and asymmetry with
environment inside the half-radius $r_{0}=$3.5 kpc. For smaller
galaxies (open circles), the average half-radius is $r_{0}=$1.8 kpc.
The absolute magnitude in $J$ is the magnitude inside $r_{0}$
\label{fig3}}
\end{figure*}

\begin{figure*}
\epsscale{0.7} \plotone{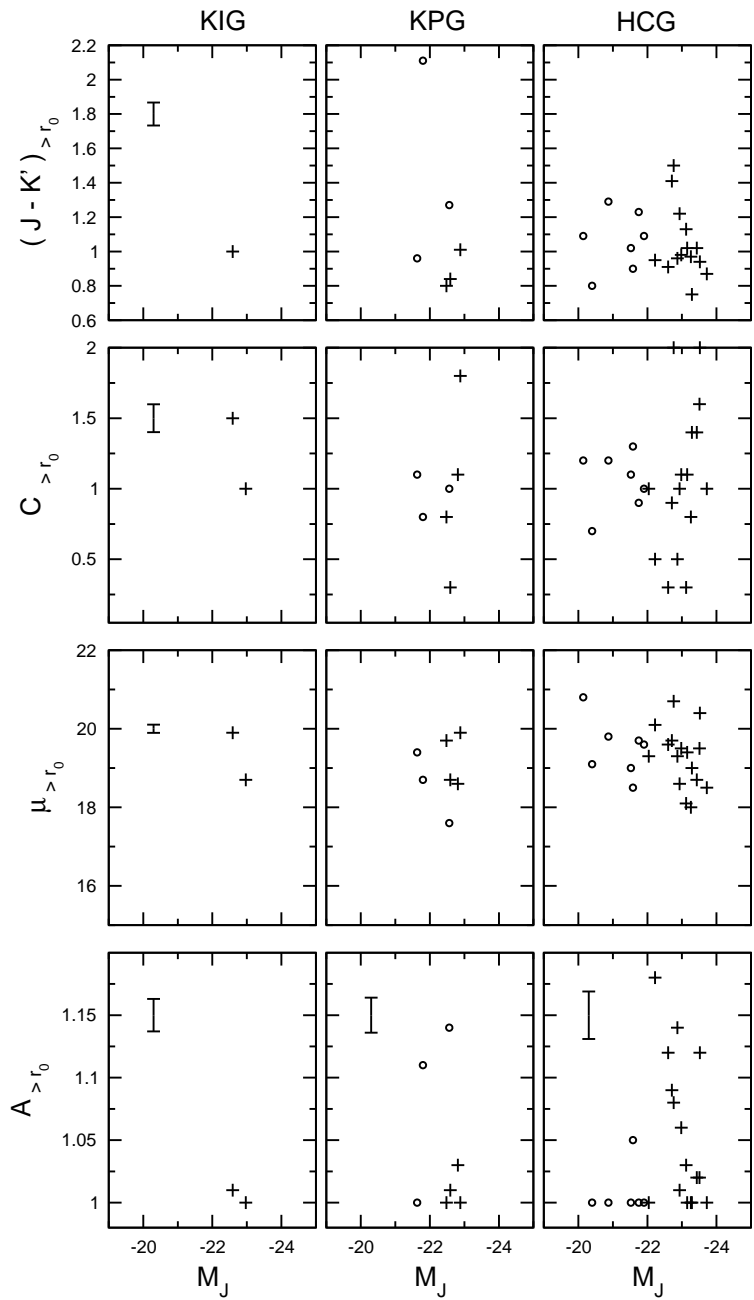}
\caption{ Variations in Early-type galaxies of isophotal parameters and asymmetry
with environment outside the half-radius $r_{0}=$3.5 kpc. For smaller galaxies (open
circles), the average half-radius is $r_{0}=$1.8 kpc. The absolute magnitude in $J$ is
the magnitude inside $r_{0}$
\label{fig4}}
\end{figure*}

We present the variations of the isophotal parameters internal to
$r_0$ in Figure~\ref{fig3} and external to $r_0$ in
Figure~\ref{fig4}. In each graph, the $x$ axis corresponds to
the $J$-band absolute magnitude of the galaxies as estimated inside $r_{0}$. One
observes a higher number of small mass galaxies in the HCG than in
the other two samples.

In this morphology group only 2 galaxies belong to the KIGs. We have
discarded them from our statistical tests. The only statistically
significant differences encountered are that the HCGs tend to be
less concentrated and have lower surface brightness inside $r_0$ 
than the KPG galaxies (see Table~\ref{tabl9}). This seems consistent
with what we observed in the optical. However, since here we are
observing in the NIR, the interpretation in terms of mass
distribution is clearer. It seems that the orbits of the stars have
higher energy in the HCGs than in the KPGs. In Paper~I, we suggest
that this effect was due to interactions. Thus, this difference would 
suggest more interactions in the HCGs than in the KPGs.

For all the other parameters we found no statistically significative
differences (see Table~\ref{tabl9}). In general, we observe much
less differences in the NIR than in the optical. Note however that
some galaxies in the HCG and KPG samples are slightly more asymmetric than
in the KIG one (see bottom panel in Figure~\ref{fig4}).

\begin{figure*}
\epsscale{0.7} \plotone{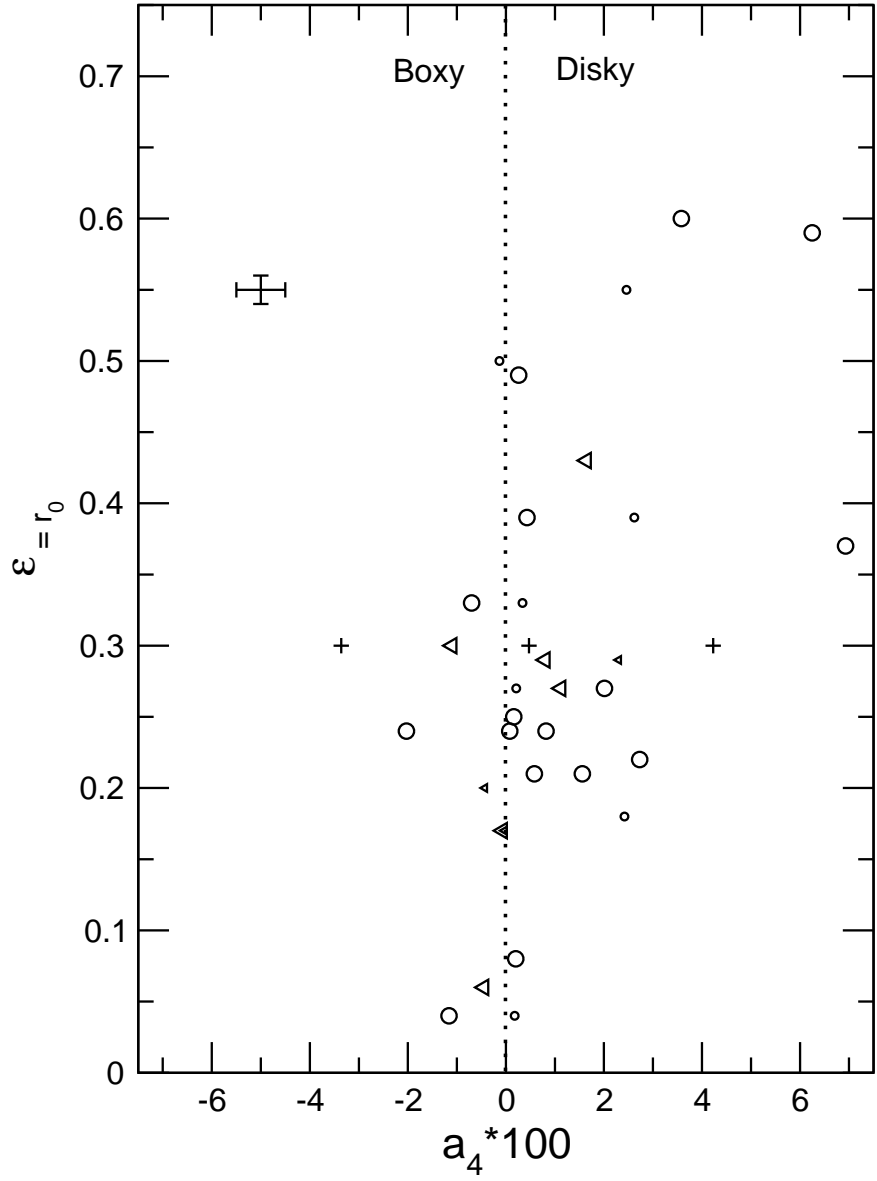}
\caption{Isophotal shape based on the $a_{4}$ parameter versus ellipticity, for the
early type galaxies. Both values were measured at $r_{0}$. The circles are for the HCG, the
 triangles for the KPG, and the plus signs for the KIG galaxies. Smaller symbols correspond to small
size galaxies (with $r_{0}=1.8$ kpc)
\label{fig5}}
\end{figure*}

As a product of the isophotal study, we have determined the isophotal 
shape $a_4$  and twist $\theta$ of the early-type galaxies. In 
Figure~\ref{fig5} we see that most of the galaxies present disky 
isophotes ($a_4>0$): 19 out of 23 (83\%) for the HCGs and 4 out of 9 (44\%) 
of the KPGs. This confirms the trends observed in the optical. 
The fractions of disky galaxies are also consistent with what we found 
in the optical.

\begin{figure*}
\epsscale{0.5} \plotone{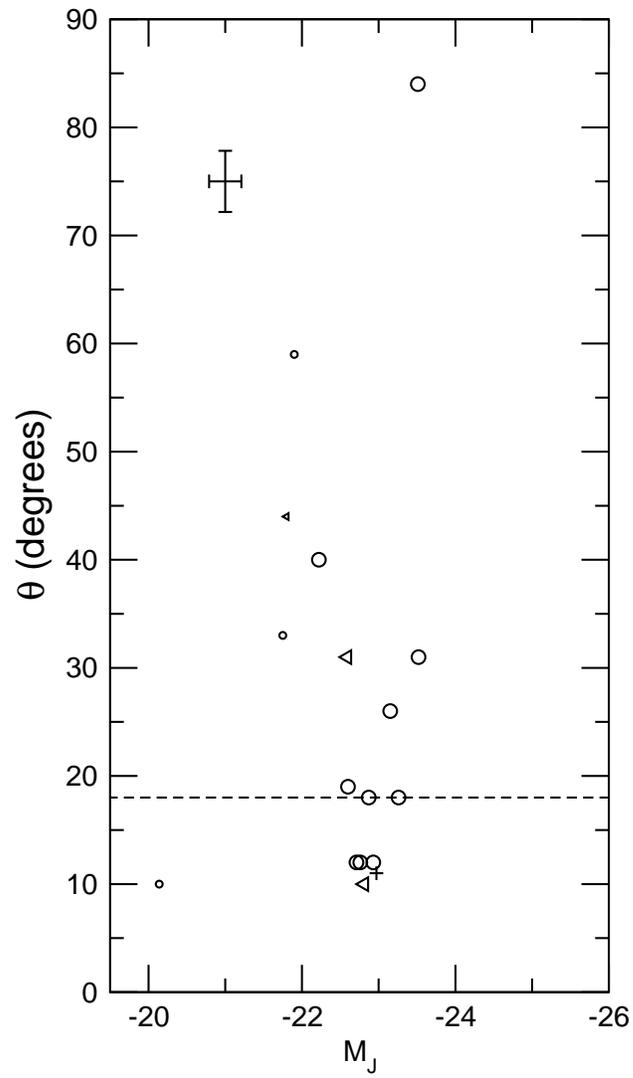}
\caption{ Early type galaxies with
twists $\theta$ as a function of the absolute magnitude in
J. Symbols are the same as in Figure~\ref{fig5}.
\label{fig6}}
\end{figure*}

Also consistent with the optical, we found the ellipticity in these
galaxies to be quite high. In Figure~\ref{fig6}, we show the twists
as a function of the absolute magnitudes in $J$. The fraction of
galaxies with large twists is in agreement with that found in the
optical: 57\% (13/23) of the HCGs, with a median $\theta$
value of 22$^\circ$, and 33\% (3/9) of the KPGs, with a median
$\theta$ value of 21$^\circ$.

\begin{figure*}
\epsscale{0.9} \plotone{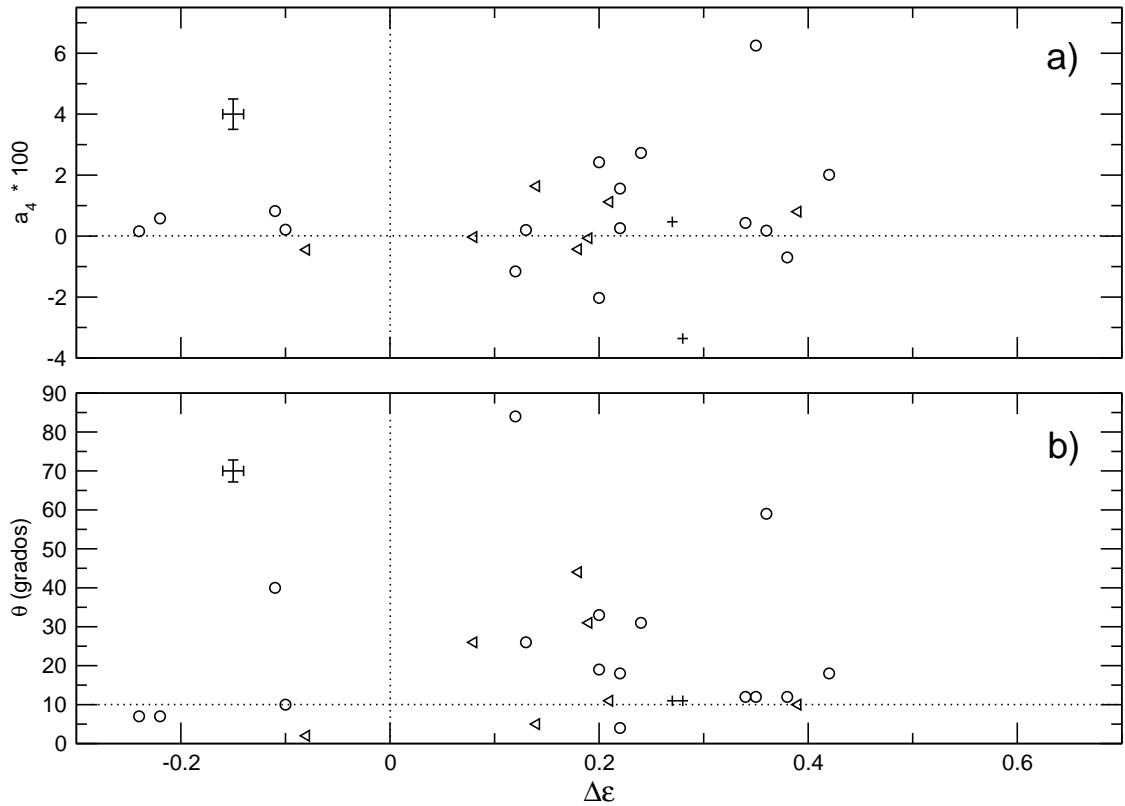}
\caption{ Ellipticity variation ($\Delta \epsilon$) versus: a) isophotal shape $a_4$ and b)
twist $\theta$ for the early type galaxies. Symbols are the same as in Figure~\ref{fig5}.
\label{fig7}}
\end{figure*}

In Figure~\ref{fig7}, we show how the isophotal para maters $a_4$ and
twists $\theta$ vary with the  difference ellipticity $\Delta
\epsilon = \epsilon_{max} - \epsilon_{min}$. A positive value of
$\Delta \epsilon$ indicates that a galaxy is rounder in its center
than at the periphery. Most of the galaxies of our samples have such
characteristic. Large values of $\Delta \epsilon$ together with
large $a_4$ and $\theta$ suggest that galaxies were affected by
interactions. Once again, the results are similar to those
observed in the optical, while no other significant difference 
is observed between the HCG and the KPG samples.

\subsubsection{Intermediate-type (Sa--Sb) galaxies}

\begin{figure*}
\epsscale{0.7} \plotone{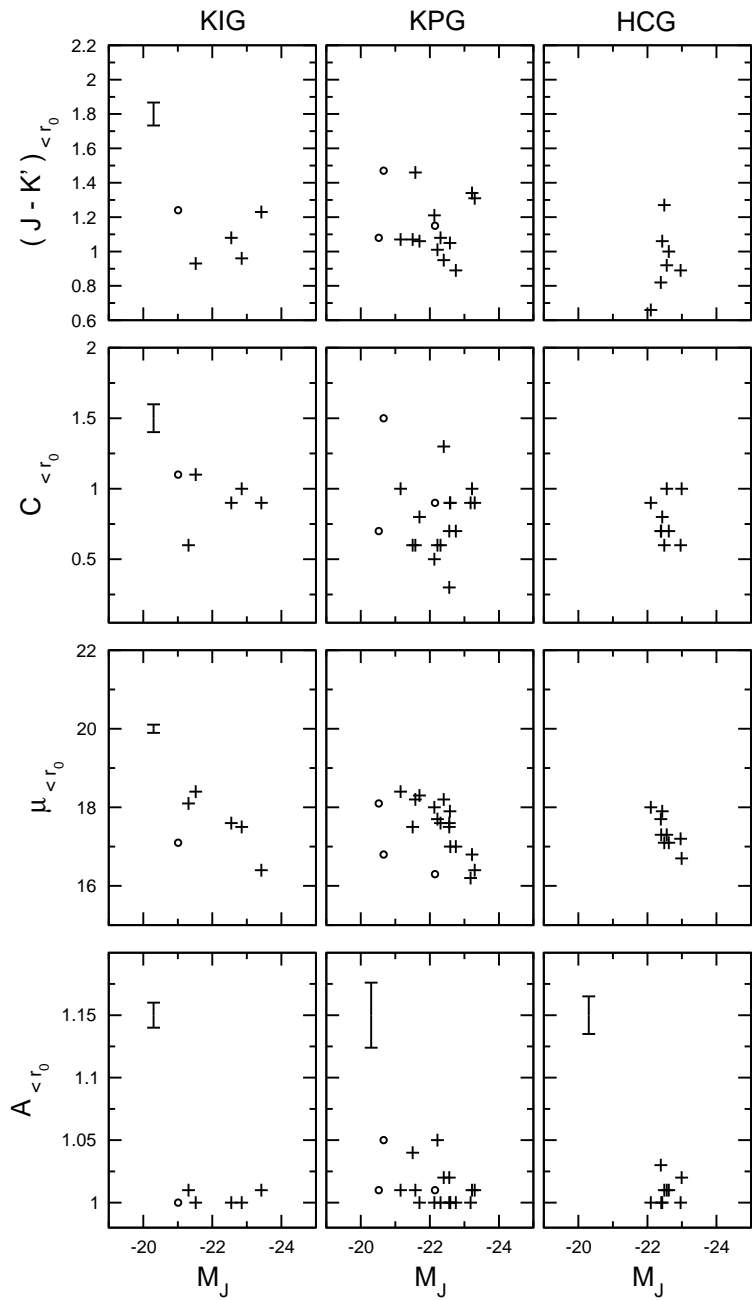}
\caption{ Variations in Intermediate-type galaxies of isophotal parameters and asymmetry
with environment inside the half-radius $r_{0}=$3.5 kpc. For smaller galaxies (open
circles), the average half-radius is $r_{0}=$1.8 kpc. The absolute magnitude in $J$ is
the magnitude inside $r_{0}$.
\label{fig8}}
\end{figure*}

\begin{figure*}
\epsscale{0.7} \plotone{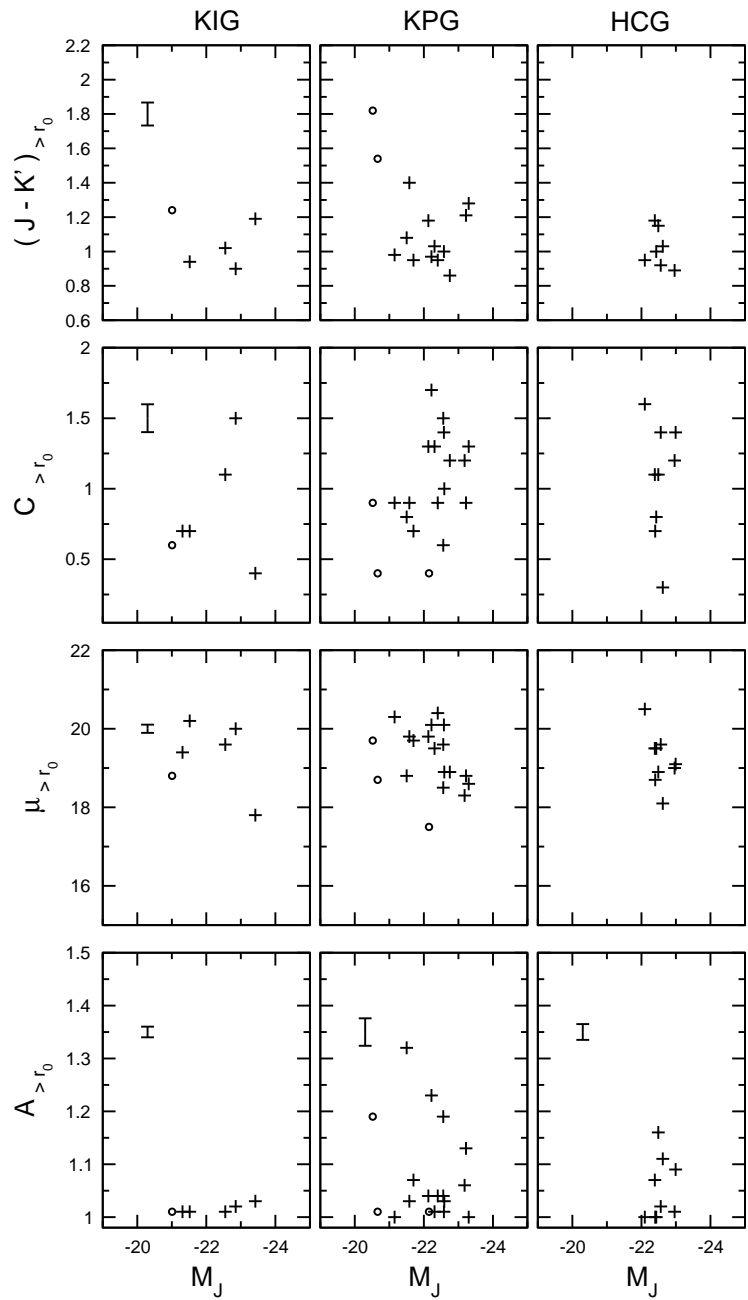}
\caption{ Variations in Intermediate-type galaxies of isophotal parameters and asymmetry
with environment outside the half-radius $r_{0}=$3.5 kpc. For smaller galaxies (open
circles), the average half-radius is $r_{0}=$1.8 kpc. The absolute magnitude in $J$ is
the magnitude inside $r_{0}$.
\label{fig9}}
\end{figure*}

In Figures~\ref{fig8} and \ref{fig9}, we show for the Sa--Sb group the
variations of the isophotal parameters internal and external to
$r_0$, respectively. The only statistically significant differences
found  are that the KPG galaxies tend to be slightly redder in their 
centers than the HCGs and more asymmetric than the KIGs outside $r_0$ 
(see Table~\ref{tabl10}). In the optical, differences between KPGs 
and HCGs were not significative. The difference in color in the NIR 
is puzzling. This cannot be due to extinction, because then we would 
have expected a difference also in the optical. Usually redder colors
suggest older stellar populations. However, it may also suggest a
difference in terms of star formation or AGN activity. For example,
the presence of numerous supergiant stars due to a recent burst
could also produce redder colors \citep{frogel87}. Alternatively, 
it could be that the KPGs are redder because of more intense AGN activity 
\citep{kotilainen94} in some of these galaxies. Both effects would 
be consistent with interaction effects.

\subsubsection{Late-type (Sbc--Im) galaxies}

\begin{figure*}
\epsscale{0.7} \plotone{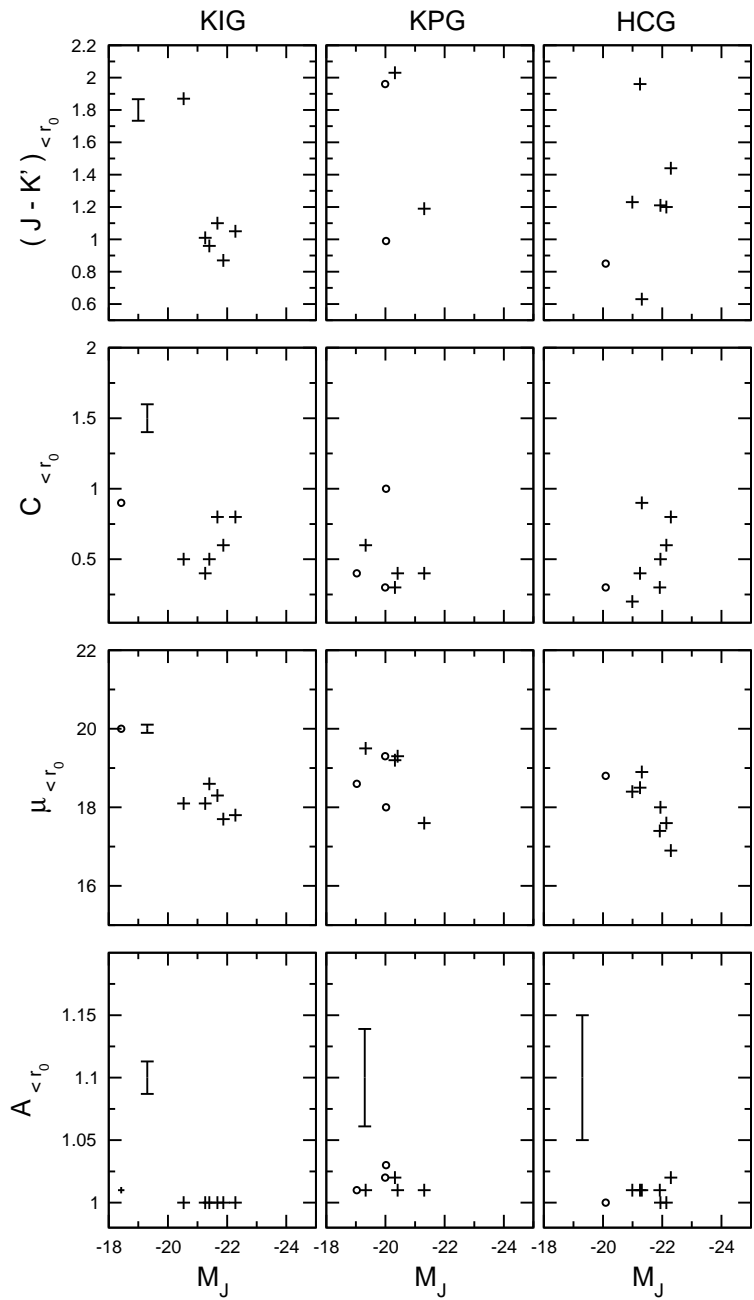}
\caption{ Variations in Late-type galaxies of Isophotal parameters and asymmetry
with environment inside the half-radius $r_{0}=$3.5 kpc. For smaller galaxies (open
circles), the average half-radius is $r_{0}=$1.8 kpc. The absolute magnitude in $J$ is
the magnitude inside $r_{0}$
\label{fig10}}
\end{figure*}

\begin{figure*}
\epsscale{0.7} \plotone{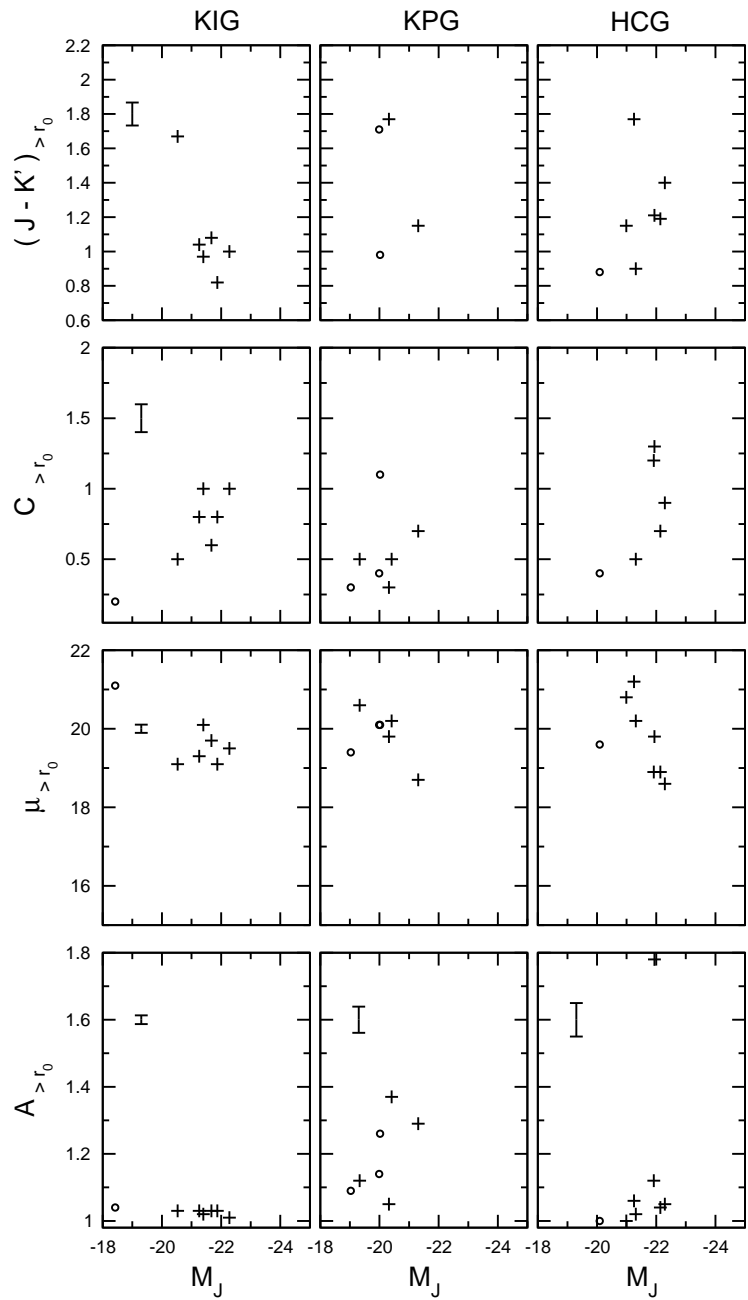}
\caption{ Variations in Late-type galaxies of Isophotal parameters and asymmetry
with environment outside the half-radius $r_{0}=$3.5 kpc. For smaller galaxies (open
circles), the average half-radius is $r_{0}=$1.8 kpc. The absolute magnitude in $J$ is
the magnitude inside $r_{0}$
\label{fig11}}
\end{figure*}

In Figure~\ref{fig10} and \ref{fig11}, we show the variations in the
Late-type galaxies of the isophotal parameters internal and external
to $r_0$, respectively. In this morphological group we observe much 
more differences than in the other two groups (Table~\ref{tabl11}). 
The KPGs tend to be fainter in the $J$-band than the HCGs and KIGs. 
The HCGs also have slightly higher surface brightness inside $r_0$ and 
are more concentrated outside this radius than the KPGs. These differences
suggest variations in stellar populations and distributions. The
lower luminosity for the KPGs and higher surface brightness for the
HCGs suggest a larger number of old stars in the nucleus of in HCGs
than that of KPGs. The higher concentration outside the radius for the HCGs
is consistent with higher energy orbits.

In this morphology group, the KPGs and HCGs are also much more
asymmetric than the KIG galaxies inside $r_0$, while outside this
radius the most asymmetric are the KPGs (see Table~\ref{tabl11}).
This suggests that this difference is related to interaction effects, 
and not due to differences in internal processes, like a bar structure, 
stochastic star formation propagation or wave density.

\subsection{Origin of the asymmetries in the NIR}

In general, we found less differences among  NIR
properties of galaxies in different environments than that
observed in the optical. This is as expected based on the
sensitivity of the different filter-bands to different stellar
populations. The NIR follows the distribution of the less massive,
but dominant, stellar populations. Since these stars formed earlier,
we expect their spatial distribution has already reached some
sort of equilibrium  within the potential well of the galaxies,
and this is independent of galaxy environment.

On the other hand, we saw that in general, and as it already observed in
the optical, the level of asymmetry appears slightly higher in the
KPGs and the HCGs than in the KIGs, being more obvious in the KPGs than
in the HCGs. However, this phenomenon in the NIR also seems to depend
on the morphology: the early and intermediate type galaxies are more
symmetric than the late-type ones. To isolate the effect of
morphology on the asymmetry we must identify clearly the nature of
the asymmetries in the NIR. Our method is similar to the one
developed in the optical. It consists in classifying the galaxies
according to different asymmetry types (see Paper~I), with the
difference that, since the NIR images are not sensible to dust
extinction, no galaxies are classified as type~2.

\begin{figure*}
\epsscale{1} \plotone{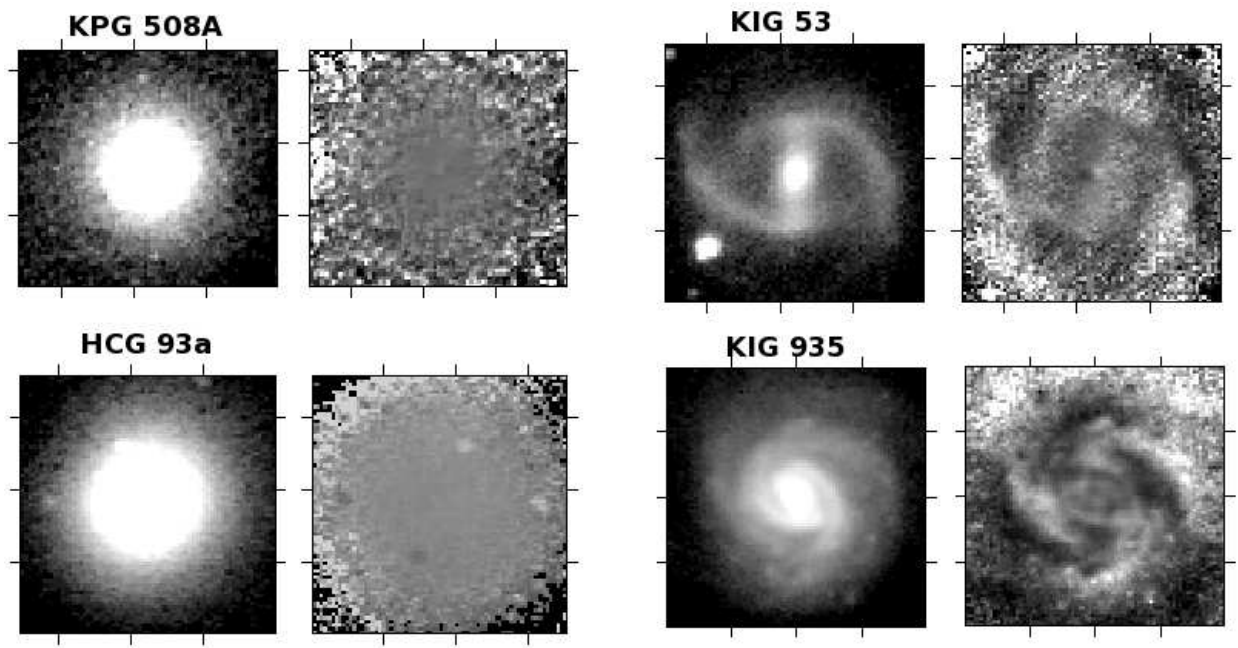}
\caption{Examples of galaxies with
type~1 asymmetry. The $J$-band images are displayed on logarithmic
scales together with their residual images. ({\it Left}): symmetric
galaxies;({\it Right}): galaxies where asymmetries are
intrinsic, related to spiral arm structures.
\label{fig12}}
\end{figure*}

\begin{figure*}
\epsscale{1} \plotone{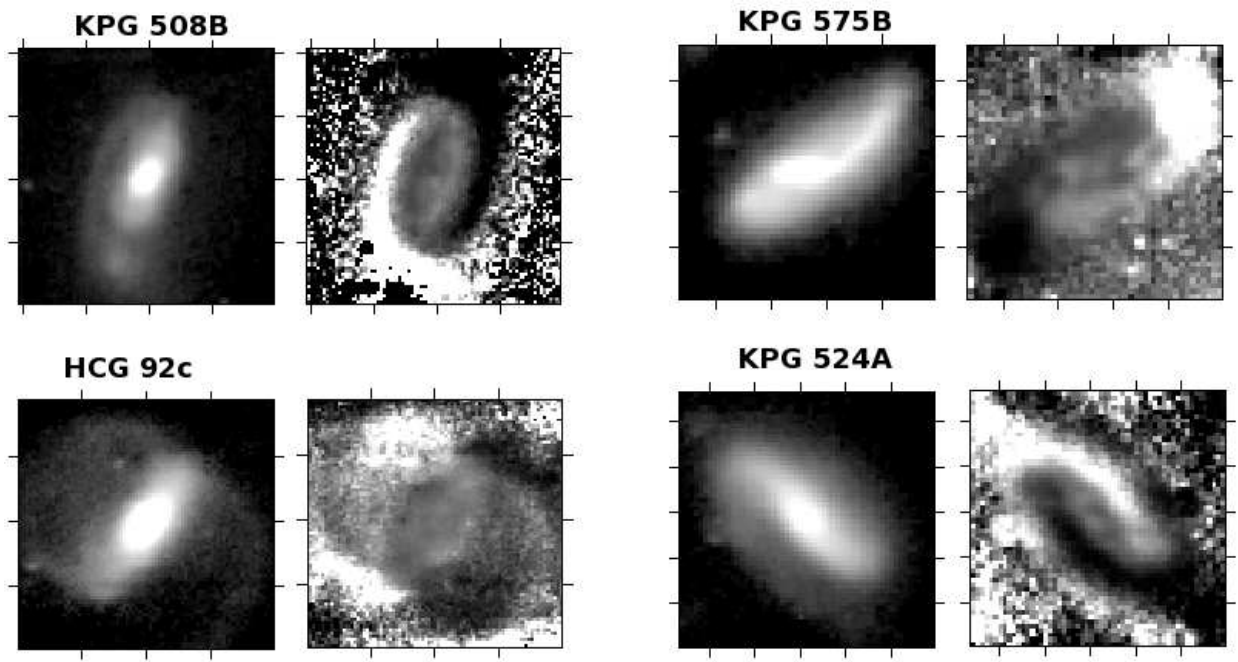}
\caption{Examples of galaxies with
type~3 asymmetry. The $J$-band images are displayed on logarithmic
scales together with their residual images. These are obvious cases
of asymmetries related to galaxy interactions.
\label{fig13}}
\end{figure*}

\begin{figure*}
\epsscale{1} \plotone{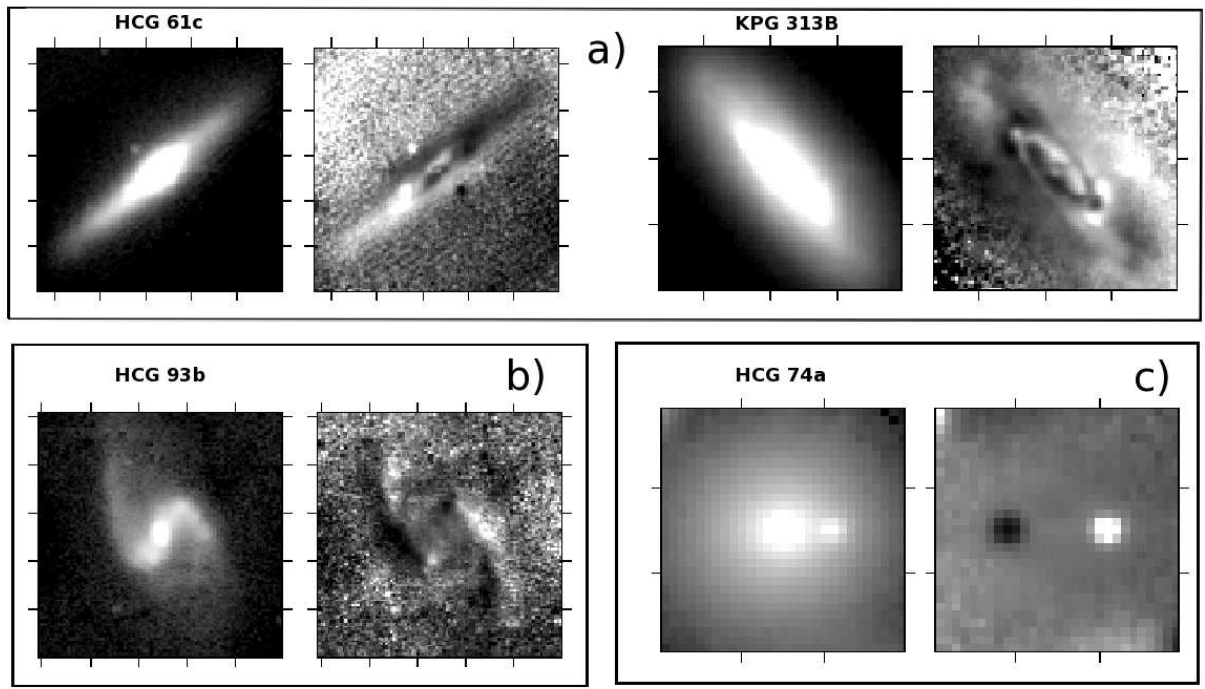}
\caption{ Examples of: a) galaxies
with type~4 asymmetry -- cause is not obvious; b) type~5 --
companion appear near center; c) type~6 -- possible double nucleus.
The $J$-band images are displayed on logarithmic scales together with
their residual images.
\label{fig14}}
\end{figure*}

In type~1, we have put all the ``symmetric'' galaxies or galaxies
with ``intrinsic'' asymmetries, related to star formation clumps
and/or spiral arms. Examples of galaxies with a type~1 asymmetry are
shown in Figure~\ref{fig12}.  In type~3 we see the most obvious
evidence of galaxy interactions, under the form of tidal tails,
plumes, connecting bridges or common envelop between galaxies.
Examples of galaxies with a type~3 asymmetry are presented in
Figure~\ref{fig13}. In type~4 we have put galaxies which are highly
asymmetric, but where the cause is not obvious. In type~5 we have
regrouped the cases where the asymmetry may be due to a smaller mass
satellite galaxy. Finally, in type~6 we have regrouped the cases
where the asymmetry is accompanied by a possible double nucleus.
Galaxies with type~4 can be found in Figure~\ref{fig14}a), those with
type~5 in Figure~\ref{fig14}b) and those with type~6 can be found in
Figure~\ref{fig14}c).

\begin{figure*}
\epsscale{1} \plotone{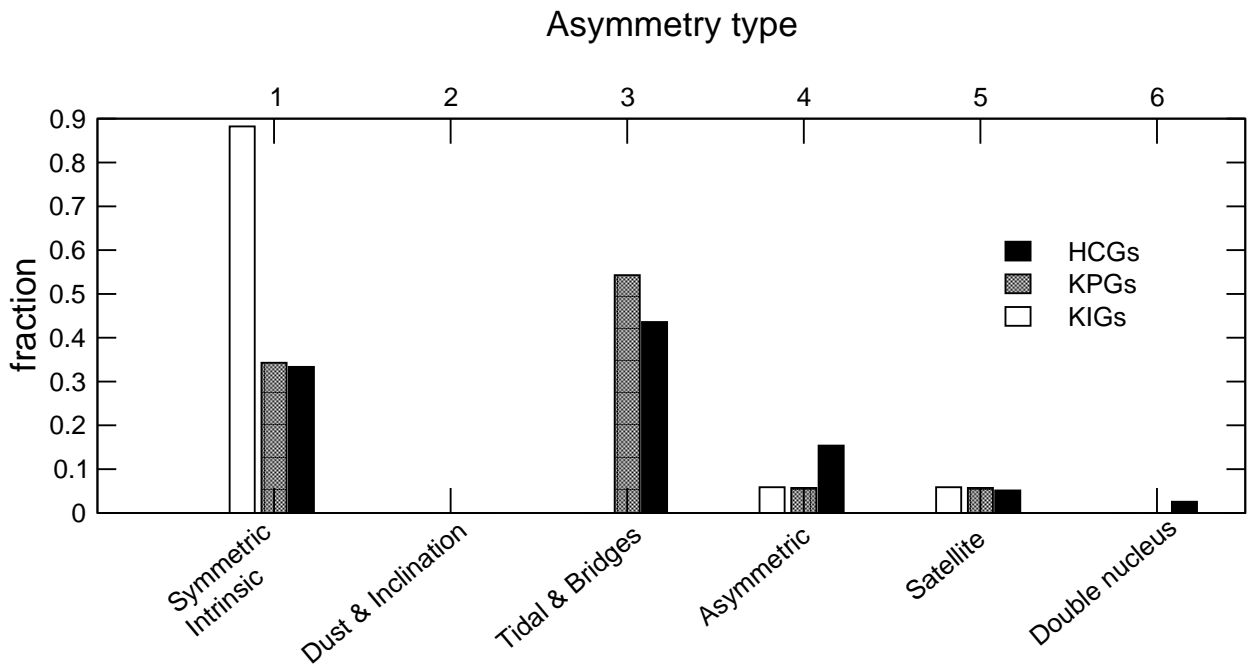}
\caption{Distribution of asymmetry types in different environments.
\label{fig15}}
\end{figure*}

The distribution of asymmetry types in the different samples, as
found in the NIR, is presented in Figure~\ref{fig15}. In the KIG
sample, 88\% of the galaxies are of type~1, the rest (12\%) are
classified as type~4 and type~5. Consequently, the fraction of 
symmetric galaxies in NIR  is higher than what it is found in the 
optical.

As in the optical, the KPGs present a higher fraction (66\%) of
asymmetries related to interactions: 54\% are of type~3, 6\% in
types 4 and 5. The rest, 34\% of KPGs, are symmetric. For the HCGs, the
number of asymmetric galaxies is also quite high, reaching 67\% of
the galaxies of our sample: 44\% are type~3, 15\% type~4, 5\%
type~5, and 3\% type~6.

In Table~\ref{tabl12}, we  compare the fraction of galaxies
belonging to each asymmetry type in the optical and NIR. We
distinguish the same trends. Comparing the classification galaxy by
galaxy, in general the asymmetry type is the same in both bands. The
most notable difference is related to dust which does not affect the
NIR images. Tidal tails and bridges seem as frequent in the NIR as
in the optical. However, we do observe a higher fraction of asymmetric
galaxies in the NIR than in the optical for the HCGs. Since we are
not seeing this effect in the KPGs, this clearly states that something is
different in the HCGs. Possibly the interactions in the HCGs are at a
more advanced stage and are affecting the oldest stellar population. Or 
possibly, interactions are now occurring  in the absence of the 
usual evidences of star formation or nuclear activity. This last 
possibility is consistent with the dry merger hypothesis.

\subsection{Color gradients and blue cores}

In normal early-type galaxies, color gradients make a galaxy core 
redder than the periphery (i.e. the gradient is negative; Peletier
et al. 1990). These color gradients can be explained, in part by the
concentration of older stellar population towards the center of the
galaxies, and in other part, by an increase in stellar metallicity 
\citep{hinkley01}. Galaxy formation models suggest that if an
elliptical galaxy form rapidly by monolithic collapse and is
undisturbed by interaction, a negative color gradients will form and
stay unchanged for most of its life-time. However, some elliptical
galaxies are known to present color gradients which are flat or
positive, i.e., bluer to the inner part \citep{michard99, im01, yang06}. 
For these galaxies, models suggest that such features in color 
gradients can be the result of mergers or past interactions 
with gas-rich galaxies.

As in Coziol \& Plauchu-Frayn (2007), we have search for NIR blue
cores in the early-type galaxies of our sample. The $J - K'$ color
gradient is defined as $\Delta (J-K')/log(r)$. According to this
definition, galaxies with blue cores have $\Delta (J-K')/log(r)<0$.
For 29 early-type galaxies in the three samples we were able to
estimate this gradient. In these galaxies we found colors consistent
with blue cores or flat gradients in 10 out of 22 (45\%) HCGs and 4
out of 6 (67\%) KPGs. The only early-type galaxy in the KIG sample, where we
were able to estimate this gradient, has $\Delta (J-K')/log(r)>0$. In
Figure~\ref{fig16}, we plot the $\Delta (J-K')/log(r)$ gradients for
 early-type galaxies of our sample. The evidence for blue cores
in early-type galaxies seems slightly higher in the KPGs than in the
HCGs. This suggests slightly older ages for the interaction events in
the CGs.

\section{Evidence of star formation produced by interactions}

In the HCGs, we have found more evidence of asymmetries in the NIR
than in the optical. This last phenomenon is not observed in the
KPGs. One possible interpretation is that in the KPGs the interactions 
involve a lot of gas, while in the HCGs these are occurring under dry
conditions. To verify this assumption, we have search for evidence
of star formation induced by interaction in the different samples.
In CGs the star formation activity was already found to be low 
(Coziol et al. 1998; 2000; 2004; Mart\'{\i}nez et al. 2010).

Using SDSS spectra together with the
STARLIGHT\footnote{http://www.starlight.ufsc.br/} spectral synthesis
code \citep{cid05}, we have classified the nuclear
activity type for 241 (26\%) galaxies in the KPG catalog with available
spectrum. In Table~\ref{tabl13} we compare the activity type in the
KPGs with those found  by Mart\'{\i}nez et al. (2010) for HCG sample. 
We clearly distinguish a larger number of emission line
galaxies  in the KPGs than in the HCGs. Also, the dominant type
of activity in the KPG galaxies is different, being the majority of them 
star forming galaxies (SFG), while those in the HCGs are low
luminosity AGNs. This comparison confirms that in the KPGs, interactions 
occur while galaxies are still rich in gas. This result
also supports our interpretation that the color difference in
terms of younger stellar populations in the KPGs is related to recent
star formation events or AGN activity. For the 19 intermediate type galaxies
in our observed sample of KPGs, we found a spectra for 7 of them. From these, 
we found that three are Sy 2 galaxies, one is LINER and two are starburst
galaxies.

The fact that we observe as much evidence of interactions in the HCGs
as in the KPGs, but no evidence of induced activity in the HCGs
compared to the KPGs confirms that the interactions in the HCGs
happen predominantly under dry conditions.

\section{Discussion}

Comparing between the optical (Paper~I) and NIR analyses, we saw
that the properties of galaxies with different morphological
types are much more similar in the NIR than in the optical and this
is independent of the environment of the galaxies. This is
consistent with older populations lying at lower energy levels in
the gravitational potential well of their galaxies. Consequently,
the  asymmetry level  induced by interactions is expected to be
higher in the optical than in the NIR. This is because it requires much
more energy to disrupt these stars.

On the other hand, asymmetric structures related to interactions
seem as frequent in the NIR as in the optical. Moreover, there
seems to be almost a one to one relation in the case of tidal tails
and bridges. This confirms our interpretation that asymmetries are related
to interactions: these correspond to real mass redistributions.

The fact that we find the HCGs to be less compact than in other
environments is, consequently, quite revealing. This observation
suggests the orbits of the stars in the HCGs are more energetic. Such 
an effect  would be achieved by increasing the number of interactions: 
the larger the number of interactions and the higher the energy of the 
stars, and consequently the more energetic, or less compact, their orbits 
in equilibrium. A galaxy in isolation, on the other hand, would be 
expected to be much more compact, which seems consistent with our 
observations for the KIGs. For the CGs, the hypothesis of multiple 
 interactions is also consistent with the numerous early-type galaxies 
found in this sample \citep{hickson88}. Our results agree with previous
works in the sense that a considerable  fractions of galaxies in CGs show 
perturbations related to interactions and/or mergers \citep{rubin91, mendes94,
verdes01}.

The question that remains for the CGs is what is the time scale of
the interaction process? Is the evolution of galaxies accelerated by
numerous interactions or are the galaxies in CGs more evolved
morphologically because they began to interact at a earlier time?
The multiple evidence of interactions in the CGs, both in the
optical and NIR, suggest these galaxies are clearly not in
equilibrium. Therefore, compact groups could not have formed that long
ago in the past.

For the pairs of galaxies, we may easily assume that the interactions are 
relatively recent. These galaxies formed in low density
environments and it took an Hubble time to two of them for meet and
interact. This interpretation is consistent with our observations.
In the KPGs, the stellar populations in the central region of the
galaxies with different morphologies seem younger, in general, than
in the HCGs. This is consistent with the spectroscopic evidence,
which shows a higher level of star formation in the KPGs compared to the
HCGs.

One key difference between the HCGs and KPGs seems to be that in the
CGs, interactions are happening under dry conditions, confirming what 
we observed in Coziol \& Plauchu-Frayn (2007). The fact that
we do not see such phenomenon in the KPGs could be due to interactions 
in these systems are very recent (this is  supported by
spectroscopy). In the CGs, a first round of interactions would have
produced the numerous early-type galaxies we now observe. Possibly
when they formed, the CGs would have experienced a more active phase of
star formation (and AGN). But now that the gas is exhausted, the
galaxies in CGs, assuming merging orbits, can only interact under
dry conditions. For the KPGs, we do not know what will be their
future. Possibly those are systems with high energy orbits, that
will interact again only after an extremely long time has passed. In
the case of the CGs the evidence of dry interaction conditions would
thus be an evidence that galaxies in these systems are now in merging
orbits. Consequently, their formation cannot be that far in the past.

\section{Conclusion}

Our analysis suggests that pairs of galaxies are young structures:
the galaxies in pairs formed and spent most of their life in
relative isolation and are just beginning to interact. This behavior
would be typical of low density environments, or what is found at
the periphery of large scale structures.

On the other hand, galaxies in CGs are obviously more evolved
morphologically and have suffered more interactions. However, based
on the abundant evidence of interactions in both the NIR and the 
optical, these systems cannot be in equilibrium. In particular, the
evidence for dry interactions in CGs is consistent with the
hypothesis that galaxies are in merging orbits. Consequently,
CGs cannot be extremely old.

Cosmologically speaking the difference in formation time between
pairs and CGs may be relatively small. That is, the two phenomena
are probably typical of the formation of structures in low density
environments and consequently their respective formation represents
relatively recent events compared to the formation of larger and
more massive structures.

According to this interpretation, one would not expect systems like
local CGs to exist at high redshifts. CGs may have formed in the
past, but these would have been much more massive than what we find
today and such systems would have be expected to merge with others
to form cluster of galaxies \citep{coziol09}.

\section{ACKNOWLEDGMENTS}

\acknowledgments We thank the CATT of San Pedro M\'artir for the
observing time given on the 2.1m telescope to realize this project
and all the personel of the observatory for their support. We also 
thank an anonymous referee for important comments and suggestions.
I. P. F. acknowledges to Drs. H. Andernach and J. M. Islas-Islas for 
their valuable feedback.

This research has made use of:

   1) SAOImage DS9, developed by Smithsonian Astrophysical Observatory; 2)
TOPCAT software provided by the UK's AstroGrid Virtual Observatory 
Project, which is funded by the Science and Technology Facilities Council 
and through the EU's Framework 6 programme; 3) Data products from the
Two Micron All Sky Survey, which is a joint project of the University of 
Massachusetts and the Infrared Processing and Analysis Center/California
Institute of Technology, funded by the National Aeronautics and Space 
Administration and the National Science Foundation; 4) Funding for the SDSS 
and SDSS-II has been provided by the Alfred P. Sloan Foundation, the Participating 
Institutions, the National Science Foundation, the U.S. Department of Energy, 
the National Aeronautics and Space Administration, the Japanese Monbukagakusho, 
the Max Planck Society, and the Higher Education Funding Council for England. 
The SDSS Web Site is http://www.sdss.org/. The SDSS is managed by the Astrophysical 
Research Consortium for the Participating Institutions. The Participating 
Institutions are the American Museum of Natural History, Astrophysical Institute 
Potsdam, University of Basel, University of Cambridge, Case Western Reserve 
University, University of Chicago, Drexel University, Fermilab, the Institute for 
Advanced Study, the Japan Participation Group, Johns Hopkins University, the Joint 
Institute for Nuclear Astrophysics, the Kavli Institute for Particle Astrophysics 
and Cosmology, the Korean Scientist Group, the Chinese Academy of Sciences (LAMOST),  
Los Alamos National Laboratory, the Max-Planck-Institute for Astronomy (MPIA), 
New Mexico State University, Ohio State University, University of Pittsburgh, 
University of Portsmouth, Princeton University, the United States Naval Observatory, 
and the University of Washington; and 5) HyperLeda database (http://leda.univ-lyon1.fr).

\clearpage

\begin{deluxetable}{lrrrlrr}

\tabletypesize{\scriptsize}
\tablecaption{Properties of the Observed KIG Galaxies}
\tablewidth{0pt}
\tablehead{
\colhead{Name} & \colhead{ R.A.}      &     \colhead{ Dec.}      & \colhead{$v_{vir}$}         &  \colhead{Morph.}  &  \colhead{$t_{J}$} &  \colhead{$t_{K'}$}     \\
\colhead{    } &  \colhead{(J$2000$)} &     \colhead{(J$2000$)}  & \colhead{(km\ s$^{-1}$)}    &  \colhead{Type}    &  \colhead{(sec.)}  &  \colhead{(sec.)}       \\
\colhead{(1)}  & \colhead{(2)}        & \colhead{(3)}            & \colhead{(4)}               & \colhead{(5)}        & \colhead{(6)}        & \colhead{(7)}       \\
}
\startdata
KIG~53  &   $   01\phn30\phn46  $   &   $   21\phn26\phn25  $   &   3199    &   SBbc    &   3600    &   996 \\
KIG~68  &   $   01\phn53\phn13  $   &   $   04\phn11\phn44  $   &   1686    &   SBa     &   4300    &   1110    \\
KIG~467 &   $   11\phn09\phn16  $   &   $   36\phn01\phn16  $   &   6560    &   SB0 &   4700    &   1176    \\
KIG~547 &   $   12\phn42\phn39  $   &   $   19\phn56\phn42  $   &   1084    &   Sbc*    &   4800    &   \nodata \\
KIG~550 &   $   12\phn44\phn26  $   &   $   37\phn07\phn16  $   &   7193    &   SBbc    &   4900    &   1110    \\
KIG~553 &   $   12\phn50\phn08  $   &   $   33\phn09\phn32  $   &   7273    &   SBb &   4400    &   970 \\
KIG~575 &   $   13\phn12\phn06  $   &   $   24\phn05\phn41  $   &   2761    &   Sb  &   4400    &   972 \\
KIG~653 &   $   14\phn51\phn38  $   &   $   40\phn35\phn57  $   &   5138    &   Sb  &   4200    &   1020    \\
KIG~805 &   $   17\phn23\phn47  $   &   $   26\phn29\phn11  $   &   4938    &   SBbc*   &   4500    &   1308    \\
KIG~812 &   $   17\phn32\phn39  $   &   $   16\phn24\phn05  $   &   3282    &   Sbc &   4700    &   810 \\
KIG~840 &   $   17\phn56\phn55  $   &   $   32\phn38\phn11  $   &   4978    &   SBbc    &   2900    &   1260    \\
KIG~841 &   $   17\phn59\phn14  $   &   $   45\phn53\phn14  $   &   5658    &   S0  &   2700    &   540 \\
KIG~852 &   $   18\phn26\phn57  $   &   $   56\phn05\phn15  $   &   8259    &   Sb  &   2400    &   \nodata \\
KIG~897 &   $   21\phn07\phn47  $   &   $   16\phn20\phn08  $   &   5090    &   Sa  &   4100    &   1030    \\
KIG~935 &   $   21\phn54\phn33  $   &   $   02\phn56\phn34  $   &   4024    &   SBc     &   4800    &   1100    \\
KIG~1001&   $   22\phn57\phn19  $   &   $   -01\phn02\phn56 $   &   3076    &   SBab    &   3600    &   1056    \\
KIG~1020&   $   23\phn29\phn03  $   &   $   11\phn26\phn42  $   &   3761    &   S0  &   \nodata &   1056    \\
\enddata
\tablecomments{Columns:
(1) catalog galaxy identification;
(2) right ascension from HyperLeda ($0^{h}00^{m}00^{s}$);
(3) declination from HyperLeda ($0^{\circ}00'00''$);
(4) radial velocity from HyperLeda, corrected for infall of Local Group towards Virgo;
(5) morphological type taken from Sulentic et al. (2006)-- a star indicates morphology was
determined in this work;
(6)--(7): total exposure time in  $J$ and $K'$ bands, respectively;
 }
\label{tabl1}
\end{deluxetable}

\begin{deluxetable}{lrrrlrr}
\tabletypesize{\scriptsize}
\tablecaption{Properties of the Observed KPG Galaxies}
\tablewidth{0pt}
\tablehead{
\colhead{Name} & \colhead{ R.A.}      &     \colhead{ Dec.}      & \colhead{$v_{vir}$}         &  \colhead{Morph.}  &  \colhead{$t_{J}$} &  \colhead{$t_{K'}$}     \\
\colhead{    } &  \colhead{(J$2000$)} &     \colhead{(J$2000$)}  & \colhead{(km\ s$^{-1}$)}    &  \colhead{Type}    &  \colhead{(sec.)}  &  \colhead{(sec.)}       \\
\colhead{(1)}  & \colhead{(2)}        & \colhead{(3)}            & \colhead{(4)}               & \colhead{(5)}        & \colhead{(6)}        & \colhead{(7)}       \\
}
\startdata
KPG~75A      &        $        02\phn45\phn09        $     &        $        32\phn59\phn23        $     &        5099     &        SBa      &        4600     &        \nodata           \\
KPG~75B      &        $        02\phn45\phn13        $     &        $        32\phn58\phn41        $     &        5169     &        SBb      &        4600     &        \nodata           \\
KPG~99A      &        $        04\phn30\phn39        $     &        $        00\phn39\phn43        $     &        3590     &        E        &        4300     &        570               \\
KPG~99B      &        $        04\phn30\phn43        $     &        $        00\phn39\phn53        $     &        3411     &        E        &        4300     &        570               \\
KPG~313A     &        $        11\phn58\phn34        $     &        $        42\phn44\phn02        $     &        1014     &        SBc      &        4400     &        \nodata           \\
KPG~313B     &        $        11\phn58\phn52        $     &        $        42\phn43\phn21        $     &        904      &        SBb      &        4800     &        430               \\
KPG~366B     &        $        13\phn13\phn26        $     &        $        27\phn48\phn08        $     &        6583     &        SBb      &        4300     &        \nodata           \\
KPG~397A     &        $        13\phn47\phn44        $     &        $        38\phn18\phn16        $     &        1631     &        Sc       &        5100     &        1320              \\
KPG~425A     &        $        14\phn23\phn42        $     &        $        34\phn00\phn32        $     &        4074     &        SBa      &        4900     &        924               \\
KPG~425B     &        $        14\phn23\phn46        $     &        $        34\phn01\phn01        $     &        3759     &        SBb      &        4900     &        924               \\
KPG~471A     &        $        15\phn44\phn21        $     &        $        41\phn05\phn08        $     &        9788     &        SBb      &        3000     &        792               \\
KPG~471B     &        $        15\phn44\phn27        $     &        $        41\phn07\phn11        $     &        9811     &        Sb       &        4600     &        1188              \\
KPG~480B     &        $        16\phn04\phn30        $     &        $        03\phn52\phn06        $     &        5612     &        Sa       &        3400     &        750               \\
KPG~492A     &        $        16\phn21\phn44        $     &        $        54\phn41\phn11        $     &        9879     &        S0       &        3800     &        912               \\
KPG~492B     &        $        16\phn15\phn23        $     &        $        26\phn37\phn04        $     &        9916     &        Sa       &        3700     &        852               \\
KPG~508A     &        $        17\phn19\phn14        $     &        $        48\phn58\phn49        $     &        7573     &        E        &        3800     &        1080              \\
KPG~508B     &        $        17\phn19\phn21        $     &        $        49\phn02\phn25        $     &        7444     &        SBb  pec &        4100     &        1080              \\
KPG~523A     &        $        17\phn46\phn07        $     &        $        35\phn34\phn10        $     &        6998     &        SBb      &        4600     &        1360              \\
KPG~523B     &        $        17\phn46\phn17        $     &        $        35\phn34\phn18        $     &        6979     &        SBb      &        4600     &        1360              \\
KPG~524A     &        $        17\phn46\phn27        $     &        $        30\phn42\phn17        $     &        4812     &        SBb      &        4800     &        960               \\
KPG~524B     &        $        17\phn46\phn31        $     &        $        30\phn41\phn54        $     &        4820     &        Sc       &        4800     &        960               \\
KPG~526A     &        $        17\phn55\phn59        $     &        $        18\phn20\phn17        $     &        3171     &        Sa       &        900      &        1188              \\
KPG~526B     &        $        17\phn56\phn03        $     &        $        18\phn22\phn23        $     &        3047     &        S0       &        2100     &        828               \\
KPG~537A     &        $        18\phn47\phn27        $     &        $        50\phn24\phn38        $     &        9325     &        SBa      &        3800     &        972               \\
KPG~542A     &        $        19\phn31\phn08        $     &        $        54\phn06\phn07        $     &        4106     &        Sb       &        3300     &        780               \\
KPG~542B     &        $        19\phn31\phn10        $     &        $        54\phn05\phn32        $     &        3955     &        E        &        3300     &        780               \\
KPG~548A     &        $        20\phn47\phn19        $     &        $        00\phn19\phn15        $     &        4272     &        SBb      &        2400     &        \nodata           \\
KPG~548B     &        $        20\phn47\phn24        $     &        $        00\phn18\phn03        $     &        3859     &        E        &        2400     &        \nodata           \\
KPG~554A     &        $        21\phn09\phn36        $     &        $        15\phn07\phn29        $     &        9222     &        S0       &        1600     &        540               \\
KPG~554B     &        $        21\phn09\phn38        $     &        $        15\phn09\phn01        $     &        9027     &        S0       &        1600     &        540               \\
KPG~557B     &        $        21\phn28\phn59        $     &        $        11\phn22\phn57        $     &        8634     &        Sc       &        2100     &        \nodata           \\
KPG~566A     &        $        22\phn19\phn27        $     &        $        29\phn23\phn44        $     &        4782     &        Sc       &        1600     &        1100              \\
KPG~566B     &        $        22\phn19\phn30        $     &        $        29\phn23\phn16        $     &        4654     &        Sd       &        1400     &        740               \\
KPG~575A     &        $        23\phn03\phn15        $     &        $        08\phn52\phn27        $     &        4950     &        Sa       &        3200     &        1020              \\
KPG~575B     &        $        23\phn03\phn17        $     &        $        08\phn53\phn37        $     &        4934     &        pec      &        3500     &        450               \\
\enddata
\tablecomments{Columns:
(1) catalog galaxy identification;
(2) right ascension from HyperLeda ($0^{h}00^{m}00^{s}$);
(3) declination from HyperLeda ($0^{\circ}00'00''$);
(4) radial velocity from HyperLeda, corrected for infall of Local Group towards Virgo;
(5) morphological type as determined in this work;
(6)--(7): total exposure time in  $J$ and $K'$ bands, respectively;
 }
\label{tabl2}
\end{deluxetable}

\begin{deluxetable}{lrrrlrr}
\tabletypesize{\scriptsize}
\tablecaption{Properties of the Observed HCG Galaxies}
\tablewidth{0pt}
\tablehead{
\colhead{Name} & \colhead{ R.A.}      &     \colhead{ Dec.}      & \colhead{$v_{vir}$}         &  \colhead{Morph.}  &  \colhead{$t_{J}$} &  \colhead{$t_{K'}$}     \\
\colhead{    } &  \colhead{(J$2000$)} &     \colhead{(J$2000$)}  & \colhead{(km\ s$^{-1}$)}    &  \colhead{Type}    &  \colhead{(sec.)}  &  \colhead{(sec.)}       \\
\colhead{(1)}  & \colhead{(2)}        & \colhead{(3)}            & \colhead{(4)}               & \colhead{(5)}        & \colhead{(6)}        & \colhead{(7)}       \\
}
\startdata

HCG~10a       &       $ 01\phn26\phn21  $      &       $    34\phn42\phn10  $      &       5269      &       SBb    &       2700      &       \nodata           \\
HCG~37a*      &       $ 09\phn13\phn39  $      &       $    29\phn59\phn35  $      &       6844      &       E      &       2250      &       860               \\
HCG~37b*      &       $ 09\phn13\phn33  $      &       $    30\phn00\phn00  $      &       6840      &       Sbc    &       2250      &       860               \\
HCG~40a*      &       $ 09\phn38\phn53  $      &       $    -04\phn50\phn56 $      &       6571      &       E      &       1600      &       900               \\
HCG~40b*      &       $ 09\phn38\phn54  $      &       $    -04\phn51\phn56 $      &       6785      &       S0     &       1600      &       900               \\
HCG~40c*      &       $ 09\phn38\phn53  $      &       $    -04\phn51\phn37 $      &       6833      &       SBbc   &       1600      &       900               \\
HCG~40d*      &       $ 09\phn38\phn55  $      &       $    -04\phn50\phn16 $      &       6435      &       SBa    &       1600      &       900               \\
HCG~40e*      &       $ 09\phn38\phn55  $      &       $    -04\phn51\phn29 $      &       6568      &       SBc    &       1600      &       900               \\
HCG~56a*      &       $ 11\phn32\phn46  $      &       $    52\phn56\phn27  $      &       8476      &       Sc     &       3600      &       900               \\
HCG~56b*      &       $ 11\phn32\phn40  $      &       $    52\phn57\phn01  $      &       8150      &       SB0    &       3600      &       900               \\
HCG~56c*      &       $ 11\phn32\phn36  $      &       $    52\phn56\phn51  $      &       8341      &       S0     &       3600      &       900               \\
HCG~56d*      &       $ 11\phn32\phn35  $      &       $    52\phn56\phn49  $      &       8577      &       S0     &       3600      &       900               \\
HCG~56e*      &       $ 11\phn32\phn32  $      &       $    52\phn56\phn21  $      &       8155      &       S0     &       3600      &       900             \\
HCG~61a       &       $ 12\phn12\phn18  $      &       $    29\phn10\phn45  $      &       3942      &       S0     &       3000      &       910               \\
HCG~61c       &       $ 12\phn12\phn31  $      &       $    29\phn10\phn05  $      &       4114      &       SBbc   &       3900      &       1000              \\
HCG~61d       &       $ 12\phn12\phn26  $      &       $    29\phn08\phn57  $      &       4138      &       SB0    &       4200      &       1090              \\
HCG~74a*      &       $ 15\phn19\phn24  $      &       $    20\phn53\phn46  $      &       12427     &       E      &       4400      &       1160              \\
HCG~74b*      &       $ 15\phn19\phn24  $      &       $    20\phn53\phn27  $      &       12282     &       E      &       4400      &       1160              \\
HCG~74c*      &       $ 15\phn19\phn25  $      &       $    20\phn53\phn57  $      &       12439     &       S0     &       4400      &       1160              \\
HCG~79a       &       $ 15\phn59\phn11  $      &       $    20\phn45\phn16  $      &       4469      &       E      &       3600      &       804               \\
HCG~79b       &       $ 15\phn59\phn12  $      &       $    20\phn45\phn47  $      &       4623      &       SB0    &       3600      &       804               \\
HCG~79c       &       $ 15\phn59\phn10  $      &       $    20\phn45\phn43  $      &       4323      &       SB0    &       3600      &       804               \\
HCG~82a       &       $ 16\phn28\phn22  $      &       $    32\phn50\phn58  $      &       11398     &       S0     &       3600      &       890               \\
HCG~82b       &       $ 16\phn28\phn27  $      &       $    32\phn50\phn46  $      &       10668     &       SBa    &       4000      &       870               \\
HCG~82c       &       $ 16\phn28\phn20  $      &       $    32\phn48\phn36  $      &       10316     &       Sd     &       4500      &       1188              \\
HCG~82d       &       $ 16\phn28\phn16  $      &       $    32\phn48\phn47  $      &       11906     &       Sa     &       4500      &       1188              \\
HCG~88a       &       $ 20\phn52\phn35  $      &       $    -05\phn42\phn40 $      &       5970      &       Sb     &       800       &       \nodata           \\
HCG~88b*      &       $ 20\phn52\phn29  $      &       $    -05\phn44\phn46 $      &       5946      &       Sb     &       3600      &       450               \\
HCG~88c*      &       $ 20\phn52\phn26  $      &       $    -05\phn46\phn20 $      &       6019      &       Sc     &       3600      &       200               \\
HCG~92b       &       $ 22\phn35\phn58  $      &       $    33\phn57\phn57  $      &       5925      &       SBb    &       4800      &       990               \\
HCG~92c       &       $ 22\phn36\phn03  $      &       $    33\phn58\phn31  $      &       6915      &       SBb    &       4800      &       990               \\
HCG~92d       &       $ 22\phn35\phn56  $      &       $    33\phn57\phn54  $      &       6781      &       S0     &       4800      &       990               \\
HCG~93a       &       $ 23\phn15\phn16  $      &       $    18\phn57\phn40  $      &       5215      &       E      &       1800      &       \nodata           \\
HCG~93b       &       $ 23\phn15\phn17  $      &       $    19\phn02\phn29  $      &       4747      &       SBc    &       2000      &       \nodata           \\
HCG~93c       &       $ 23\phn15\phn03  $      &       $    18\phn58\phn24  $      &       5207      &       SBa    &       1600      &       450               \\
HCG~93d       &       $ 23\phn15\phn33  $      &       $    19\phn02\phn52  $      &       5248      &       S0     &       1200      &       300               \\
HCG~94a*      &       $ 23\phn17\phn13  $      &       $    18\phn42\phn27  $      &       12113     &       E      &       1300      &       650               \\
HCG~94b*      &       $ 23\phn17\phn12  $      &       $    18\phn42\phn03  $      &       12047     &       E      &       1300      &       650             \\
HCG~98a*      &       $ 23\phn54\phn10  $      &       $    00\phn22\phn58  $      &       7835      &       SB0    &       3600      &       1440              \\
HCG~98b*      &       $ 23\phn54\phn12  $      &       $    00\phn22\phn37  $      &       7939      &       S0     &       3600      &       1440              \\
\enddata
\tablecomments{Columns: 
(1) catalog galaxy identification; 
(2) right ascension from HyperLeda ($0^{h}00^{m}00^{s}$); 
(3) declination from HyperLeda ($0^{\circ}00'00''$); 
(4) radial velocity from Hickson et al. (1992), corrected for infall of Local Group towards Virgo; 
(5) morphological type as determined in this work; 
(6)--(7): total exposure time in $J$ and $K'$ bands, respectively;
 }
\label{tabl3}
\end{deluxetable}

\begin{deluxetable}{lcccccccccc}
\tabletypesize{\scriptsize} \tablecaption{Properties of observed vs.
catalog galaxies} \tablewidth{0pt} \tablehead{
\colhead{Sample}     & \colhead{$M_{B}$}   & \colhead{$P_{MW}$}       & \colhead{$M_{K_s}$}     & \colhead{$P_{MW}$}   &\colhead{$D_{B}$}     &  \colhead{$P_{MW}$}      &\colhead{$D_{K_s}$}     &  \colhead{$P_{MW}$}    &\colhead{$V_{vir}$}     &  \colhead{$P_{MW}$}     \\
\colhead{}           & \colhead{(mag)}     & \colhead{}               & \colhead{(mag)}         & \colhead{}           &\colhead{(kpc)}       &  \colhead{}              &\colhead{(kpc)}         &  \colhead{}            &\colhead{(km s$^{-1}$)} &  \colhead{}             \\
\colhead{(1)}        & \colhead{(2)}       & \colhead{(3)}            & \colhead{(4)}           & \colhead{(5)}        &\colhead{(6)}         & \colhead{(7)}            &\colhead{(8)}           & \colhead{(9)}          & \colhead{(10)}          &  \colhead{(11)}        \\
}
\startdata
KIG                 &  -20.60/-20.30      &  \underline{0.007}         &  -23.65/-23.27         &  \underline{0.036}    &  27/22                &  0.091                 &  18/16                &  0.088                 &  5034/6296                &  \underline{0.0161}  \\
KPG                 &  -20.36/-20.32      &  0.266                     &  -23.55/-23.89         &  0.122                &  25/23                &  0.465                 &  18/16                &  0.350                 &  4942/6326                &  0.2218               \\
HCG                 &  -20.57/-20.01      &  \underline{0.001}         &  -24.60/-23.60         &  \underline{0.002}    &  31/24                &  \underline{0.001}     &  23/16                &  \underline{0.001}     &  6783/7970                &  0.2295               \\
\enddata
\tablecomments{Columns: 
(1) sample identification;
(2) median absolute magnitude in $B$, $M_{B}$ of observed/catalog galaxies; 
(3) probability $P$ for $M_{B}$; 
(4) median absolute magnitude in $K$, $M_{K}$,of observed/catalog galaxies; 
(5) probability $P$ for $M_{K}$;  
(6) median diameter in optical, $D_{B}$, of observed/catalog galaxies; 
(7) probability $P$ for $D_{B}$; 
(8) median diameter in NIR $D_{K}$ of the observed/catalog galaxies; 
(9) probability $P$ for $D_{K}$; 
(10) median redshift, $v_{vir}$, of observed/catalog galaxies; 
(11) probability $P$ for $v_{vir}$. $P$ values were obtained from Mann-Whitney tests; underlined values indicate significative
differences between the samples.} \label{tabl4}
\end{deluxetable}

\begin{deluxetable}{clccrr}
\tabletypesize{\scriptsize}
\tablecaption{Observing runs}
\tablewidth{0pt} \tablehead{
\colhead{Run}    &  \colhead{Date} & \colhead{Filters} & \colhead{Seeing} &
\colhead{$\sigma_J$} & \colhead{$\sigma_{K'}$} \\
\colhead{}       &   \colhead{}    & \colhead{}        & \colhead{(FWHM)} &
\colhead{(mag)}        & \colhead{(mag)} \\
\colhead{(1)}    & \colhead{(2)}   & \colhead{(3)}     & \colhead{(4)}    &
\colhead{(5)}    & \colhead{(6)}}
\startdata
1 &   2006 Aug          & $J, K'$ &  1.2$''$  &  $\pm0.17$ & $\pm0.08$ \\
2 &   2007 May          & $J, K'$ &  2.2$''$  &  $\pm0.08$ & $\pm0.04$ \\
3 &   2007 Sep          & $J, K'$ &  1.8$''$  &  $\pm0.06$ & $\pm0.09$ \\
4 &   2008 May          & $J, K'$ &  1.4$''$  &  $\pm0.02$ & $\pm0.03$ \\
5 &   2008 Jul          & $J, K'$ &  1.4$''$  &  $\pm0.02$ & $\pm0.08$ \\
\enddata
\tablecomments{Columns: 
(1) running number; 
(2) observation date; 
(3) average FWHM measured on standard stars used for focus; 
(4) Filters; 
(5)--(6): Calibration uncertainties for $J$ and K$'$, respectively } \label{tabl5}
\end{deluxetable}

\begin{deluxetable}{lrrrrrrl}

\tabletypesize{\scriptsize}
\tablecaption{Observed properties of KIG Galaxies}
\tablewidth{0pt}
\tablehead{
\colhead{Name}  &    \colhead{$C$}     &  \colhead{$C$}    &   \colhead{$J-K'$}   &  \colhead{$J-K'$}  &  \colhead{$A$}     &  \colhead{$A$}      &  \colhead{Asymmetry}          \\
\colhead{    }  &  \colhead{$<r_0$}    &  \colhead{$>r_0$} &  \colhead{$<r_0$}    &  \colhead{$>r_0$}  &  \colhead{$<r_0$}  &  \colhead{$>r_0$}   &  \colhead{Type}        \\
\colhead{(1)}   & \colhead{(2)}        & \colhead{(3)}     & \colhead{(4)}       & \colhead{(5)}     &  \colhead{(6)}       & \colhead{(7)}       & \colhead{(8)}            \\
}
\startdata
KIG~53  &   0.5 &   0.5 &   1.87    &   1.67    &   1.00    &   1.03    &   Intrinsic                  \\
KIG~68  &   1.1 &   0.6 &   1.24    &   1.24    &   1.00    &   1.01    &   Symmetric                  \\
KIG~467 &   0.9 &   1.5 &   1.14    &   1.00    &   1.00    &   1.01    &   Symmetric                 \\
KIG~547 &   0.9 &   0.2 &   \nodata &   \nodata &   1.01    &   1.04    &   Intrinsic           \\
KIG~550 &   0.8 &   1.0 &   1.05    &   1.00    &   1.00    &   1.01    &   Intrinsic                 \\
KIG~553 &   1.0 &   1.5 &   0.96    &   0.90    &   1.00    &   1.02    &   Satellite                 \\
KIG~575 &   1.1 &   0.7 &   0.93    &   0.94    &   1.00    &   1.01    &   Intrinsic                 \\
KIG~653 &   0.9 &   1.1 &   1.08    &   1.02    &   1.00    &   1.01    &   Intrinsic                 \\
KIG~805 &   0.5 &   1.0 &   0.96    &   0.97    &   1.00    &   1.02    &   Intrinsic                 \\
KIG~812 &   0.8 &   0.6 &   1.10    &   1.08    &   1.00    &   1.03    &   Intrinsic                 \\
KIG~840 &   0.4 &   0.8 &   1.01    &   1.04    &   1.00    &   1.03    &   Intrinsic                 \\
KIG~841 &   1.0 &   1.0 &   1.15    &   \nodata &   1.03    &   1.00    &   Symmetric              \\
KIG~852 &   0.6 &   0.7 &   \nodata &   \nodata &   1.01    &   1.01    &   Symmetric           \\
KIG~897 &   0.9 &   0.4 &   1.23    &   1.19    &   1.01    &   1.03    &   Symmetric                 \\
KIG~935 &   0.6 &   0.8 &   0.87    &   0.82    &   1.00    &   1.03    &   Intrinsic                 \\
KIG~1001 &  0.9 &   0.5 &   \nodata &   \nodata &   1.00    &   1.01    &   Symmetric          \\
KIG~1020 &  1.1 &   0.4 &   \nodata &   \nodata &   1.00    &   1.04    &   Asymmetric         \\

\enddata
\tablecomments{Columns: 
(1) catalog galaxy identification; 
(2) concentration inside $r_0$; 
(3) concentration outside $r_0$; 
(4) $J-K'$ color inside $r_0$; 
(5) $J-K'$ color outside $r_0$; 
(6) asymmetry inside $r_0$; 
(7) asymmetry outside $r_0$.; 
(8) asymmetry type.} \label{tabl6}
\end{deluxetable}

\begin{deluxetable}{lrrrrrrl}

\tabletypesize{\scriptsize}
\tablecaption{Observed properties of KPG Galaxies}
\tablewidth{0pt}
\tablehead{
\colhead{Name}  &    \colhead{$C$}     &  \colhead{$C$}    &   \colhead{$J-K'$}   &  \colhead{$J-K'$}  &  \colhead{$A$}     &  \colhead{$A$}      &  \colhead{Asymmetry}          \\
\colhead{    }  &  \colhead{$<r_0$}    &  \colhead{$>r_0$} &  \colhead{$<r_0$}    &  \colhead{$>r_0$}  &  \colhead{$<r_0$}  &  \colhead{$>r_0$}   &  \colhead{Type}        \\
\colhead{(1)}   & \colhead{(2)}        & \colhead{(3)}     & \colhead{(4)}       & \colhead{(5)}     &  \colhead{(6)}       & \colhead{(7)}       & \colhead{(8)}            \\
}
\startdata

KPG~75A      &         0.9     &        1.0     &        \nodata     &        \nodata     &        1.00     &        1.03    &     Tidal      \\
KPG~75B      &         0.3     &        0.6     &        \nodata     &        \nodata     &        1.02     &        1.19    &     Tidal      \\
KPG~99A      &         0.7     &        1.0     &        1.24        &        1.27        &        1.02     &        1.14    &     Bridges    \\
KPG~99B      &         1.1     &        0.8     &        1.76        &        2.11        &        1.00     &        1.11    &     Tidal      \\
KPG~313A     &         0.4     &        0.3     &        \nodata     &        \nodata     &        1.01     &        1.09    &     Intrinsic    \\
KPG~313B     &         1.5     &        0.4     &        1.47        &        1.54        &        1.05     &        1.01    &     Satellite    \\
KPG~366B     &         0.7     &        1.5     &        \nodata     &        \nodata     &        1.00     &        1.04    &     Tidal      \\
KPG~397A     &         1.0     &        1.1     &        0.99        &        0.98        &        1.03     &        1.26    &     Asymmetric    \\
KPG~425A     &         0.7     &        0.9     &        1.08        &        1.82        &        1.01     &        1.19    &     Tidal      \\
KPG~425B     &         1.0     &        0.9     &        1.07        &        0.98        &        1.01     &        1.00    &     Tidal      \\
KPG~471A     &         0.6     &        1.7     &        1.01        &        0.97        &        1.05     &        1.23    &     Tidal      \\
KPG~471B     &         0.5     &        1.3     &        1.21        &        1.18        &        1.00     &        1.04    &     Tidal      \\
KPG~480B     &         0.9     &        1.3     &        1.31        &        1.28        &        1.01     &        1.00    &     Symmetric    \\
KPG~492A     &         1.0     &        1.8     &        1.16        &        1.01        &        1.00     &        1.00    &     Symmetric    \\
KPG~492B     &         0.9     &        1.4     &        1.05        &        1.00        &        1.00     &        1.01    &     Symmetric    \\
KPG~508A     &         1.0     &        0.8     &        0.91        &        0.80        &        1.00     &        1.00    &     Symmetric    \\
KPG~508B     &         0.7     &        1.2     &        0.89        &        0.86        &        1.00     &        1.51    &     Tidal      \\
KPG~523A     &         1.3     &        0.9     &        0.95        &        0.95        &        1.02     &        1.04    &     Symmetric    \\
KPG~523B     &         0.6     &        1.3     &        1.08        &        1.03        &        1.00     &        1.01    &     Intrinsic    \\
KPG~524A     &         0.6     &        0.8     &        1.07        &        1.08        &        1.04     &        1.32    &     Tidal      \\
KPG~524B     &         0.6     &        0.5     &        \nodata     &        \nodata     &        1.01     &        1.12    &     Tidal      \\
KPG~526A     &         0.9     &        0.4     &        1.15        &        \nodata     &        1.01     &        1.01    &     Symmetric    \\
KPG~526B     &         1.3     &        0.3     &        0.78        &        0.84        &        1.00     &        1.01    &     Symmetric    \\
KPG~537A     &         0.6     &        0.9     &        1.46        &        1.40        &        1.01     &        1.03    &     Tidal      \\
KPG~542A     &         0.8     &        0.7     &        1.06        &        0.95        &        1.00     &        1.07    &     Bridges    \\
KPG~542B     &         1.1     &        1.1     &        0.94        &        0.96        &        1.00     &        1.00    &     Symmetric    \\
KPG~548A     &         0.9     &        1.2     &        \nodata     &        \nodata     &        1.00     &        1.06    &     Tidal      \\
KPG~548B     &         1.0     &        1.1     &        \nodata     &        \nodata     &        1.00     &        1.03    &     Tidal      \\
KPG~554A     &         0.8     &        0.7     &        \nodata     &        \nodata     &        1.00     &        1.02    &     Symmetric    \\
KPG~554B     &         0.9     &        0.6     &        \nodata     &        \nodata     &        1.01     &        1.03    &     Symmetric    \\
KPG~557B     &         0.4     &        0.5     &        \nodata     &        \nodata     &        1.01     &        1.37    &     Satellite    \\
KPG~566A     &         0.3     &        0.3     &        2.03        &        1.77        &        1.02     &        1.05    &     Tidal      \\
KPG~566B     &         0.3     &        0.4     &        1.96        &        1.71        &        1.02     &        1.14    &     Tidal      \\
KPG~575A     &         1.0     &        0.9     &        1.34        &        1.21        &        1.01     &        1.13    &     Asymmetric    \\
KPG~575B     &         0.4     &        0.7     &        1.19        &        1.15        &        1.01     &        1.29    &     Tidal      \\

\enddata
\tablecomments{ Columns are the same as defined in Table~\ref{tabl6}
}
\label{tabl7}
\end{deluxetable}

\begin{deluxetable}{lrrrrrrl}

\tabletypesize{\scriptsize}
\tablecaption{Observed properties of HCG Galaxies}
\tablewidth{0pt}
\tablehead{
\colhead{Name}  &    \colhead{$C$}     &  \colhead{$C$}    &   \colhead{$J-K'$}   &  \colhead{$J-K'$}  &  \colhead{$A$}     &  \colhead{$A$}      &  \colhead{Asymmetry}          \\
\colhead{    }  &  \colhead{$<r_0$}    &  \colhead{$>r_0$} &  \colhead{$<r_0$}    &  \colhead{$>r_0$}  &  \colhead{$<r_0$}  &  \colhead{$>r_0$}   &  \colhead{Type}        \\
\colhead{(1)}   & \colhead{(2)}        & \colhead{(3)}     & \colhead{(4)}       & \colhead{(5)}     &  \colhead{(6)}       & \colhead{(7)}       & \colhead{(8)}            \\
}
\startdata
HCG~10a       &         1.0      &       1.4      &       \nodata   &       \nodata   &       1.02      &       1.09     &      Tidal  \\
HCG~37a*      &         0.9      &       2.0      &       1.51      &       1.50      &       1.00      &       1.08     &      Asymmetric  \\
HCG~37b*      &         0.4      &       2.3      &       1.96      &       1.77      &       1.01      &       1.06     &      Asymmetric  \\
HCG~40a*      &         1.1      &       2.0      &       0.91      &       0.94      &       1.00      &       1.12    &      Tidal   \\
HCG~40b*      &         1.1      &       0.3      &       0.91      &       0.91      &       1.00      &       1.00    &      Symmetric   \\
HCG~40c*      &         0.6      &       0.7      &       1.20      &       1.19      &       1.00      &       1.04    &      Asymmetric   \\
HCG~40d*      &         0.8      &       0.8      &       1.06      &       1.00      &       1.00      &       1.00    &      Symmetric   \\
HCG~40e*      &         0.3      &       0.4      &       0.85      &       0.88      &       1.00      &       1.00    &      Tidal   \\
HCG~56a*      &         0.2      &       2.1      &       1.23      &       1.15      &       1.01      &       1.00     &      Asymmetric  \\
HCG~56b*      &         0.8      &       0.9      &       1.46      &       1.41      &       0.92      &       1.09     &      Tidal  \\
HCG~56c*      &         0.4      &       1.0      &       1.08      &       1.09      &       1.00      &       1.00     &      Bridges  \\
HCG~56d*      &         0.3      &       1.2      &       1.36      &       1.29      &       1.00      &       1.00     &      Bridges  \\
HCG~56e*      &         0.4      &       1.2      &       1.12      &       1.09      &       \nodata   &       \nodata  &      \nodata\\
HCG~61a       &         1.1      &       1.0      &       1.17      &       1.22      &       1.01      &       1.01     &      Symmetric  \\
HCG~61c       &         0.8      &       0.9      &       1.44      &       1.40      &       1.02      &       1.05     &      Asymmetric  \\
HCG~61d       &         1.0      &       1.0      &       0.86      &       \nodata   &       1.00      &       1.00     &      Symmetric  \\
HCG~74a*      &         0.3      &       1.4      &       1.13      &       1.02      &       1.01      &       1.02     &      Double nuclei  \\
HCG~74b*      &         0.6      &       1.1      &       1.02      &       0.98      &       1.00      &       1.06     &      Bridges  \\
HCG~74c*      &         0.7      &       0.5      &       1.00      &       0.95      &       0.96      &       1.18     &      Bridges  \\
HCG~79a       &         0.8      &       1.1      &       1.05      &       1.02      &       1.00      &       1.00     &      Symmetric  \\
HCG~79b       &         0.6      &       1.3      &       1.03      &       0.90      &       1.03      &       1.05     &      Tidal  \\
HCG~79c       &         0.6      &       0.7      &       0.76      &       0.80      &       1.01      &       1.00     &      Bridges  \\
HCG~82a       &         0.7      &       1.4      &       0.76      &       0.75      &       1.00      &       1.00     &      Symmetric  \\
HCG~82b       &         0.6      &       1.2      &       0.89      &       0.89      &       1.00      &       1.01     &      Symmetric  \\
HCG~82c       &         0.5      &       1.3      &       1.21      &       1.21      &       1.00      &       1.78     &      Tidal  \\
HCG~82d       &         0.9      &       1.6      &       0.66      &       0.95      &       1.00      &       1.00     &      Symmetric  \\
HCG~88a       &         0.7      &       0.7      &       \nodata   &       \nodata   &       1.00      &       1.00    &      Asymmetric   \\
HCG~88b*      &         0.7      &       1.1      &       0.82      &       1.18      &       1.03      &       1.07    &      Tidal \&   satellite   \\
HCG~88c*      &         0.9      &       0.5      &       0.63      &       0.90      &       1.01      &       1.02    &      Intrinsic   \\
HCG~92b       &         0.7      &       0.3      &       1.00      &       1.03      &       1.01      &       1.11     &      Bridges  \\
HCG~92c       &         0.6      &       1.1      &       1.27      &       1.15      &       1.01      &       1.16     &      Tidal  \\
HCG~92d       &         0.9      &       0.3      &       1.10      &       1.13      &       1.00      &       1.03     &      Bridges  \\
HCG~93a       &         1.1      &       1.6      &       \nodata   &       \nodata   &       1.00      &       1.02     &      Symmetric  \\
HCG~93b       &         0.3      &       1.2      &       \nodata   &       \nodata   &       1.01      &       1.12     &      Satellite  \\
HCG~93c       &         1.0      &       1.4      &       0.92      &       0.92      &       1.01      &       1.02     &      Symmetric  \\
HCG~93d       &         1.3      &       0.9      &       1.04      &       1.23      &       1.00      &       1.00     &      Symmetric  \\
HCG~94a*      &         0.6      &       1.0      &       0.81      &       0.87      &       1.00      &       1.00     &      Symmetric  \\
HCG~94b*      &         0.6      &       1.1      &       0.80      &       1.02      &       \nodata   &       \nodata  &      Symmetric\\
HCG~98a*      &         0.6      &       0.8      &       0.99      &       0.97      &       1.00      &       1.00     &      Bridges  \\
HCG~98b*      &         1.0      &       0.5      &       0.98      &       0.96      &       1.00      &       1.14     &      Bridges  \\
\enddata
\tablecomments{ Columns are the same as defined in Table~\ref{tabl6}
}
\label{tabl8}
\end{deluxetable}

\begin{deluxetable}{lrrr}
\tabletypesize{\scriptsize}
\tablewidth{0pt}
\tablecaption{Median values of the properties of Early-type galaxies in different environments}
\tablehead{
\colhead{Property}  &  \colhead{HCGs}   &   \colhead{KPGs} & \colhead{$P_{MW}$}  \\
\colhead{(1)}       &  \colhead{(2)}    &   \colhead{(3)}  & \colhead{(4)}
}
\startdata

$M_{J}$             & -22.8             & -22.6            & 0.2023                   \\
$M_{K'}$            & -23.9             & -23.8            & 0.4164                   \\
$\mu_{<r_{0}}$      & 17.4              & 17.1             & \underline{0.0473}       \\
$\mu_{>r_{0}}$      & 19.4              & 18.7             & 0.2019                   \\
$(J-K')_{<r_{0}}$   & 1.03              & 1.05             & 0.3372                   \\
$(J-K')_{>r_{0}}$   & 1.02              & 0.99             & 0.4420                   \\
$C_{<r_{0}}$        & 0.7               & 1.0              & \underline{0.0201}       \\
$C_{>r_{0}}$        & 1.0               & 0.8              & 0.1510                   \\
$A_{<r_{0}}$        & 1.00              & 1.00             & 0.3825                   \\
$A_{>r_{0}}$        & 1.02              & 1.02             & 0.4371                   \\

\enddata
\tablecomments{Columns: 
(1) properties compared in each sample: absolute magnitude in $J$ (mag) inside $r_{0}$,
absolute magnitude in $K'$ (mag) inside $r_{0}$, surface brightness in $J$ (mag arcsec$^{-2}$) inside and outside $r_{0}$,  $J-K'$ color
(mag) inside and outside $r_{0}$, concentration index inside and
outside $r_{0}$, asymmetry level inside and outside $r_{0}$; 
(2) and (3) medians of galaxy properties in HCGs and KPGs, respectively; 
(4) $P$ values obtained from Mann-Whitney tests; underlined values values
indicate significative differences.} \label{tabl9}
\end{deluxetable}

\begin{deluxetable}{lrrrrrr}
\tabletypesize{\scriptsize}
\tablewidth{0pt}
\tablecaption{Median values of the properties of Intermediate-type galaxies in different environments}
\tablehead{
\colhead{Property}  &  \colhead{HCGs}  &  \colhead{KPGs}  & \colhead{KIGs} &  \colhead{HCG-KPG} & \colhead{HCG-KIG}  & \colhead{KPG-KIG} \\
\colhead{(1)}       & \colhead{(2)}    &  \colhead{(3)}   & \colhead{(4)}  &  \colhead{(5)}     & \colhead{(6)}      & \colhead{(7)}
}
\startdata

$M_{J}$             &  -22.5    &  -22.3   &   -22.0   & 0.1450               & 0.2643     & 0.4620                 \\
$M_{K'}$            &  -23.5    &  -23.1   &   -23.1   & 0.0563               & 0.2226     & 0.4845                 \\
$\mu_{<r_{0}}$      &   17.3    &   17.6   &    17.6   & 0.3200               & 0.2971     & 0.4241                 \\
$\mu_{>r_{0}}$      &   19.1    &   19.6   &    19.5   & 0.4027               & 0.3616     & 0.4746                 \\
$(J-K')_{<r_{0}}$   &   0.92    &   1.08   &    1.08   & \underline{0.0100}   & 0.1010     & 0.3632                 \\
$(J-K')_{>r_{0}}$   &   1.00    &   1.06   &    1.20   & 0.1158               & 0.3194     & 0.2436                \\
$C_{<r_{0}}$        &   0.7     &   0.8    &    0.9    & 0.4900               & 0.3616     & 0.4746                 \\
$C_{>r_{0}}$        &   1.1     &   0.9    &    0.7    & 0.3651               & 0.2970     & 0.2082                 \\
$A_{<r_{0}}$        &   1.01    &   1.01   &    1.00   & 0.3970               & 0.1175     & 0.0658                 \\
$A_{>r_{0}}$        &   1.02    &   1.04   &    1.01   & 0.1674               & 0.3313     & \underline{0.0255}     \\

\enddata
\tablecomments{Columns: 
(1) properties compared in each sample: absolute magnitude in $J$ (mag) inside $r_{0}$,
absolute magnitude in $K'$ (mag) inside $r_{0}$, average values of the surface brightness 
in $J$ (mag arcsec$^{-2}$) inside and outside $r_{0}$, average $J-K'$ color (magnitude) 
inside and outside $r_{0}$, concentration index inside and outside $r_{0}$, average asymmetry 
level inside and outside $r_{0}$;  
(2)--(4): medians of galaxy properties in each sample: HCG, KPG, and KIG, respectively; 
(5)--(7): $P$ values obtained with Dunn's post-tests; 
underlined value indicates a significative difference between the samples.} \label{tabl10}
\end{deluxetable}

\begin{deluxetable}{lrrrrrr}
\tabletypesize{\scriptsize}
\tablewidth{0pt}
\tablecaption{Median values of the properties of Late-type galaxies in different environments}
\tablehead{
\colhead{Property}  &  \colhead{HCGs}  &  \colhead{KPGs}  & \colhead{KIGs} &  \colhead{HCG-KPG} & \colhead{HCG-KIG}  & \colhead{KPG-KIG} \\
\colhead{(1)}       & \colhead{(2)}    &  \colhead{(3)}   & \colhead{(4)}  &  \colhead{(5)}     & \colhead{(6)}      & \colhead{(7)}
}
\startdata

$M_{J}$             &    -21.6  &  -20.0    &   -21.4    & \underline{0.0054} & 0.3063               & \underline{0.0265}     \\
$M_{K'}$            &    -23.0  &  -22.4    &   -22.3    & 0.1576             & 0.2669               & 0.5000                 \\
$\mu_{<r_{0}}$      &     18.2  &   19.2    &    18.1    & \underline{0.0362} & 0.3472               & 0.2026                 \\
$\mu_{>r_{0}}$      &     19.7  &   20.1    &    19.5    & 0.4309             & 0.4869               & 0.2206                 \\
$(J-K')_{<r_{0}}$   &     1.21  &   1.58    &    1.03    & 0.2849             & 0.2669               & 0.0857                 \\
$(J-K')_{>r_{0}}$   &     1.19  &   1.43    &    1.02    & 0.3175             & 0.1830               & 0.0857                 \\
$C_{<r_{0}}$        &     0.5   &   0.4     &    0.6     & 0.4765             & 0.1210               & 0.0601                 \\
$C_{>r_{0}}$        &     1.1   &   0.5     &    0.8     & \underline{0.0209} & 0.1234               & 0.1520                 \\
$A_{<r_{0}}$        &     1.01  &   1.01    &    1.00    & \underline{0.0332} & \underline{0.0356}   & \underline{0.0046}     \\
$A_{>r_{0}}$        &     1.06  &   1.20    &    1.05    & \underline{0.0278} & 0.1921               & \underline{0.0003}     \\

\enddata
\tablecomments{ Columns are the same as defined in Table~\ref{tabl10}}
\label{tabl11}
\end{deluxetable}

\begin{deluxetable}{lccc}
\tabletypesize{\scriptsize}
\tablecaption{Distribution of asymmetry types}
\tablewidth{0pt} \tablehead{
\colhead{Asymmetry}     &  \colhead{HCG}      &   \colhead{KPG}     &   \colhead{KIG}      \\
\colhead{Type}          &  \colhead{OPT/NIR}  &   \colhead{OPT/NIR} &   \colhead{OPT/NIR}  \\
\colhead{}              &  \colhead{\% / \%}    &   \colhead{\% / \%}  &   \colhead{\% / \%}    \\
\colhead{(1)}    & \colhead{(2)}   & \colhead{(3)}     & \colhead{(4)}   }
\startdata
1  Symmetric            & $44/33$  &  $27/34$  &  $60/88$  \\
2  Dust  \& Inclination & $12/0 $  &  $4 /0 $  &  $ 8/0 $  \\
3  Tidal \& Bridges     & $31/44$  &  $52/54$  &  $ 0/0 $  \\
4  Asymmetric           & $6/15 $  &  $8 /6 $  &  $19/6 $  \\
5  Satellite            & $6/5  $  &  $6 /6 $  &  $13/6 $  \\
6  Double nucleus       & $1/3  $  &  $3 /0 $  &  $ 0/0 $  \\
\enddata
\tablecomments{Columns: 
(1) asymmetry type; 
(2)--(4): distribution of asymmetry types in the optical and in this
NIR, for the HCG, KPG, and KIG galaxies, respectively. }
\label{tabl12}
\end{deluxetable}

\begin{deluxetable}{lccccccccccc}
\tabletypesize{\scriptsize} \tablewidth{0pt}
\tablecaption{Distribution of nuclear activity type in the HCG and
KPG} \tablehead{ \colhead{Sample}  & \colhead{number} & \colhead{No
emission} & \colhead{Emission}  & \colhead{SFG}     & \colhead{TO} &
\colhead{AGN}     &
\\
\colhead{(1)}     & \colhead{(2)}    & \colhead{(3)}        & \colhead{(4)}     &
\colhead{(5)}     & \colhead{(6)}    & \colhead{(7)}    
}
\startdata
HCG     &  270 &   100 (37\%) & 170 (63\%) &  54  (32\%)   &  39 (23\%)  &  77 (45\%)  \\
KPG     &  241 &    43 (18\%) & 198 (82\%) &  111 (56\%)   &  35 (18\%)  &  52 (26\%)  \\

\enddata
\tablecomments{Columns: 
(1) sample identification; 
(2) number of galaxies with available spectra; 
(3) fraction of galaxies with no emission lines; 
(4) fraction of galaxies showing emission lines; 
(5) fraction of star forming galaxies; 
(6) fraction of transition objects; 
(7) fraction of  AGN (Sy2, LINER, and Sy1). Data for the HCG galaxies has been obtained from Mart\'inez et al. (2010)}
\label{tabl13}
\end{deluxetable}

\clearpage

\onecolumn

\clearpage

\clearpage

\clearpage

\clearpage

\clearpage

\clearpage

\clearpage

\clearpage

\clearpage

\clearpage

\clearpage

\clearpage

\clearpage

\clearpage

\clearpage

\begin{figure*}
\epsscale{0.5} \plotone{f16.eps}
\caption{Color gradients (in magnitude) as a function of absolute  magnitudes in $J$
inside $3.5$ kpc in the early-type galaxies: HCG ({\it circles}), KPG ({\it triangles}),
KIG ({\it plus signs}).
\label{fig16}}
\end{figure*}

\clearpage

\end{document}